\documentclass[authoryear,preprint,review,11pt]{elsarticle}
\usepackage[top=2.5cm,bottom=3cm,width=16.5cm]{geometry}
\usepackage{lineno}
\usepackage{longtable}
\usepackage{graphicx, subfigure}
\usepackage{amssymb}
\usepackage{epsfig}
\usepackage[authoryear]{natbib}
\usepackage{lscape}
\usepackage{morefloats}
\usepackage{url}
\usepackage{hyperref}
\hypersetup{
    colorlinks=true,
    linkcolor=red,
    filecolor=magenta,      
    urlcolor=cyan,
}

\usepackage{textcomp}
\usepackage{lscape}

\renewcommand{\vec}[1]{{\mathbfit #1}}

\newcommand{\xx}{ \vec x}

\newcommand{\aap}{    {\it Astron. Astrophys.}}

\newcommand{\apj}{    {\it Astrophys. J.}}
\newcommand{\apjl}{   {\it Astrophys. J. Lett.}}

\newcommand{\grl}{    {\it Geophys. Res. Lett.}}

\newcommand{\jgr}{    {\it J. Geophys. Res.}}

\newcommand{\solphys}{{\it Solar Phys.}}
 
\newcommand{\ssr}{    {\it Space Sci. Rev.}} 
 
\chardef\us=`\_

\journal{asr}
\begin{document}
\begin{frontmatter}
\author{A.C. Umuhire\fnref{label1}}
\ead{angelaciany@gmail.com}
\title{Trends and Characteristics of High-Frequency Type II Bursts Detected by CALLISTO Spectrometers}
  \author{J. Uwamahoro\fnref{label2}}
  \author{K. Sasikumar Raja\fnref{label3}}
  \author{A. Kumari\fnref{label4}}
  \author[label5]{C. Monstein}
    
\address[label1]{University of Rwanda, College of Science and Technology, Kigali, Rwanda\\(\textit{angelaciany@gmail.com})}
\address[label2]{University of Rwanda, College of Education, Rwanda\\(\textit{mahorojpacis@gmail.com})}
\address[label3]{Indian Institute of Astrophysics, II Block, Koramangala, Bengaluru - 560 034, India \\(\textit{sasikumar.raja@iiap.res.in})}
\address[label4]{Department of Physics, University of Helsinki, P.O. Box 64, FI-00014 Helsinki, Finland\\(\textit{anshusingh628@gmail.com})}
\address[label5]{Istituto Ricerche Solari (IRSOL), Università della Svizzera italiana (USI), CH-6605 Locarno-Monti, Switzerland\\(\textit{christian.monstein@irsol.usi.ch})}

\begin{abstract}
\noindent
Solar radio type II bursts serve as early indicators of incoming geo-effective space weather events such as coronal mass ejections (CMEs). In order to investigate the origin of high-frequency type II bursts (HF type II bursts), we have identified 51 of them (among 180 type II bursts from SWPC reports) that are observed by ground-based Compound Astronomical Low-cost Low-frequency Instrument for Spectroscopy and Transportable Observatory (CALLISTO) spectrometers and whose upper-frequency cutoff (of either fundamental or harmonic emission) lies in between 150 MHz-450 MHz during 2010 - 2019. We found that 60\% of HF type II bursts, whose upper-frequency cutoff $\geq$ 300 MHz originate from the western longitudes. Further, our study finds a good correlation ($\sim 0.73$) between the average shock speed derived from the radio dynamic spectra and the corresponding speed from CME data. Also, we found that analyzed HF type II bursts are associated with wide and fast CMEs located near the solar disk. In addition, we have analyzed the spatio-temporal characteristics of two of these high-frequency type II bursts and compared the derived from radio observations with those derived from multi-spacecraft CME observations from SOHO/LASCO and STEREO coronagraphs. 

\end{abstract}


\end{frontmatter}

\section{Introduction}
\noindent   Solar radio bursts (SRBs) in metric wavelengths, can be classified as i) type I bursts, which are usually associated with active regions \citep{elgaroy1977solar, Ram2013, Mercier2015}; type II bursts, which are generally related with coronal shocks \citep{cane1984type, kumari2017b}; iii) type III bursts, which are often associated with flares \citep{Fainberg1970, Ratcliffe2014}; iv) type IV bursts, generally associated with CMEs and flux ropes \citep{Gary1985,Ramesh2004,Sas2014,Car2017, morosan2020electron, Kumari2021}; and v) type V bursts, which often accompanies type III bursts \citep{stewart1965solar, Suzuki1985}. 
\noindent
Type II radio bursts are slowly drifting and long-lasting (few minutes) in the solar dynamic spectra. They are believed to be plasma emission produced by a shock when it propagates through the solar corona \citep{Mann1995,Zlotnik1998, Cliver2004,Schmidt}. These bursts are often associated with the shock propagating ahead of a CME in the solar corona \citep[e.g.,][]{Gopalswamy1998, kumari2017a, 2021A&A...647L..12M, Majumdar2021}. Type II bursts can sometimes occur without any CME, and are then associated with flares \citep{Magdaleni2012, Su_2015}. Meter-wavelength type II bursts associated with CMEs can be used as a proxy to estimate the near-Sun kinematics and dynamics of a CME \citep{shanmugaraju2017heights, kumari2017a, kumari2017c}. Type II bursts can also be observed in the decameter/hectometer (DH) frequency range ($\leq 15 $ MHz). There are various properties of meter/DH-wavelength type II radio burst, e.g. drift rates, frequency range, duration, etc., which can be used to study various properties of the `middle' and/or `upper' corona. While the meter-wavelength type II is of particular interest for CMEs near the Sun, the DH type IIs can indicate the presence of interplanetary CMEs \citep[ICMEs; e.g.,][]{Mujiber2012, Vasanth2013}. Hence, type II bursts can also act as an indicator of geomagnetic storms. \\

\noindent The starting frequency of the type II burst is often below 150 MHz, and its drift rate is $\sim$ 0.3 MHz/s \citep{Maxwell, Mann1995}. The high starting frequency implies a radio source closer to the solar disk, hence crucial for space weather forecasting purposes \citep{Makela}.  Several studies discussed the origin of type II bursts using different methods. For example, \cite{Gopalswamy09} used the leading edge method to show that the  CME height at type II onset is similar to the height at the minimum of the Alfv\'{e}n speed  \citep{Gopal2001}.  \cite{Ramesh} analyzed 41 type II bursts observed by the Gauribidanur Radioheliograph (GRAPH) at the Gauribidanur observatory and found that 92\% of the bursts were located at or above the associated CME leading edge.  \cite{Cho} used the 2011 February 13 high starting frequency type II burst to check if the associated CME generated it. They found that a CME driven shock can explain the origin of a type II burst in the low corona. Some other studies, e.g. \cite{Cane1988} reported that almost all HF type II bursts are due to flares as their onset times often coincide with impulsive phases of flares. \cite{Vrsnak1995} reported about the time coincidence between the back-extrapolated onset time of type II bursts,  with the peak energy release of the 28 high-frequency ($>$ 273 MHz) type II bursts.  The idea was also supported by \cite{Shanmugaraju2009} who did an exclusive study on the source of the high ($\geq$ 100 MHz) , low  ($\leq$ 50 MHz) and middle (50 MHz $\leq$f $\leq$ 100 MHz) starting frequency type II bursts. \cite{Gopalswamy} argued that all type II bursts are due to CMEs regardless of the wavelength.\\

\noindent 
Among these SRBs, the type II and type IV  bursts are of special interest for space weather monitoring as they often occur in associations with coronal mass ejections (CMEs), solar flares (SFs), and solar energetic particles (SEPs) \citep{Gosling1976, Macqueen, Gopalswamy1998, kumari2019direct, Ndacyayisenga_2021}. CMEs, SFs, and SEPs are usually considered major space weather drivers \citep{Cane1988, Reiner, Gopalswamy2010a, Car2020}.\\

\noindent
SEP events are due to the accelerated ions originating in large solar eruptions  such as CMEs and solar flares \citep{Gopalswamy2003}. It is known that type II radio bursts are produced by electrons accelerated in MHD shocks. This  implies a high degree of association between type II bursts and SEP events \citep{Kahler1984,Cane1990}. Previous studies found a close relationship and association between DH type II bursts, SEP events with fast and wide CMEs \cite{Gopalswamy2002,Gopalswamy2003b}. The similar analysis by \cite{Lara2003} in the metric wavelength range (25 MHz to 625 MHz) confirmed that the trend extends to higher frequencies.\\

\noindent 
It is worth mentioning that after the launch of the WIND/WAVES \citep{Bougeret1995}, several similar studies that deal with the low-frequency SRBs observed by the radio receiver band-2 (RAD2) in the frequency range $\approx 1-14$ MHz have been carried out \citep{Bale1999,Gopalswamy2004a, Gopalswamy2004b,Krupar2019, Gopalswamy2019}. Note that such observations overlap with the field-of-view (FOV) of the Solar and Heliospheric Observatory (SOHO) mission's Large Angle and Spectrometric Coronagraphs \citep[LASCO;][]{Brueckner1995}. \\

\noindent
The recent analysis by \cite{Umuhire2021} considered higher starting frequency to check whether the universal drift rate-frequency relationship extends to higher frequencies. They analyzed 128 type II radio bursts and emphasized on the 40 higher-starting-frequency bursts. The study compared two populations, type II bursts with the starting frequency less than 150 MHz and those with starting frequency greater or equal to 150 MHz. The radio burst data were obtained from different ground spectrometers, covering the period from 2010 to 2016.\\

\noindent
Our current study is a statistical analysis of 51 type II bursts whose upper-frequency cut-off lies in between 150 MHz and 450 MHz, observed by a network of the Compound Astronomical Low-cost Low Frequency for Transportable Observatory (e-Callisto), aims in analyzing the trend of type II bursts in general and provides a detailed investigation on the origin of the analyzed high-frequency type II bursts. The data used are from 2010-2019. We were motivated to use the CALLISTO ground based spectrometer data as a locally installed instrument, to check its capacity in providing radio burst data compared to other ground based spectrometers. In this article, bandwidth, duration, starting frequency, and drift rate of type II bursts are measured, to investigate their association with CMEs and/or solar flares.\\

\noindent In section \ref{sect.2} and section \ref{sect.3}, we discussed about the observational data and method, respectively. In section \ref{sect.4}, we present the data analysis and results. The summary and conclusions of the article is described in section \ref{sect.5}.\\
\section{Observations}
\label{sect.2}

\subsection{Radio observations}   
 \noindent
The present study executes a statistical analysis of 51 high starting frequency (150 MHz-450 MHz) type II bursts observed in the metric wavelengths using data obtained from the e-Callisto network to examine their origin. The sample of events was observed for nine years in SC 24 (from 2010 to 2019).
To begin with, we have identified the Type II bursts using the National Oceanic and Atmospheric Administration (NOAA) solar and geophysical reports\footnote{\url{ ftp://ftp.ngdc.noaa.gov/STP/swpc_products/}}. From that list, we further identified the HF type II observed by different CALLISTO stations.
\label{sec2.1}
\noindent
For radio data, we used the spectral data obtained using various observatories of the e-Callisto network \footnote{\url{http://www.e-callisto.org/}}, including the CALLISTO station in Rwanda\footnote{\url{http://www.e-callisto.org/StatusReports/status_83_V01.pdf}}.
\noindent
The CALLISTO  spectrometer is a new concept for solar radio observations, designed, developed, and distributed by Christian Monstein \citep{Benz}. CALLISTO is a low-frequency radio spectrometer used to monitor metric and decametric radio bursts and has been installed at different longitudes, enabling monitoring the radio emissions $24 \times 7$. All CALLISTO spectrometers constitute the e-Callisto network for data transfer online. More than 152 stations are in operation, and 55 of them provide real-time data to the server (one frame every 15 minutes) located at the University of Applied Sciences (FHNW) in Brugg/Windisch, Switzerland.\\

\begin{figure}
\centering
\begin{tabular}{ll}
{\includegraphics[width=9cm,height=8cm]{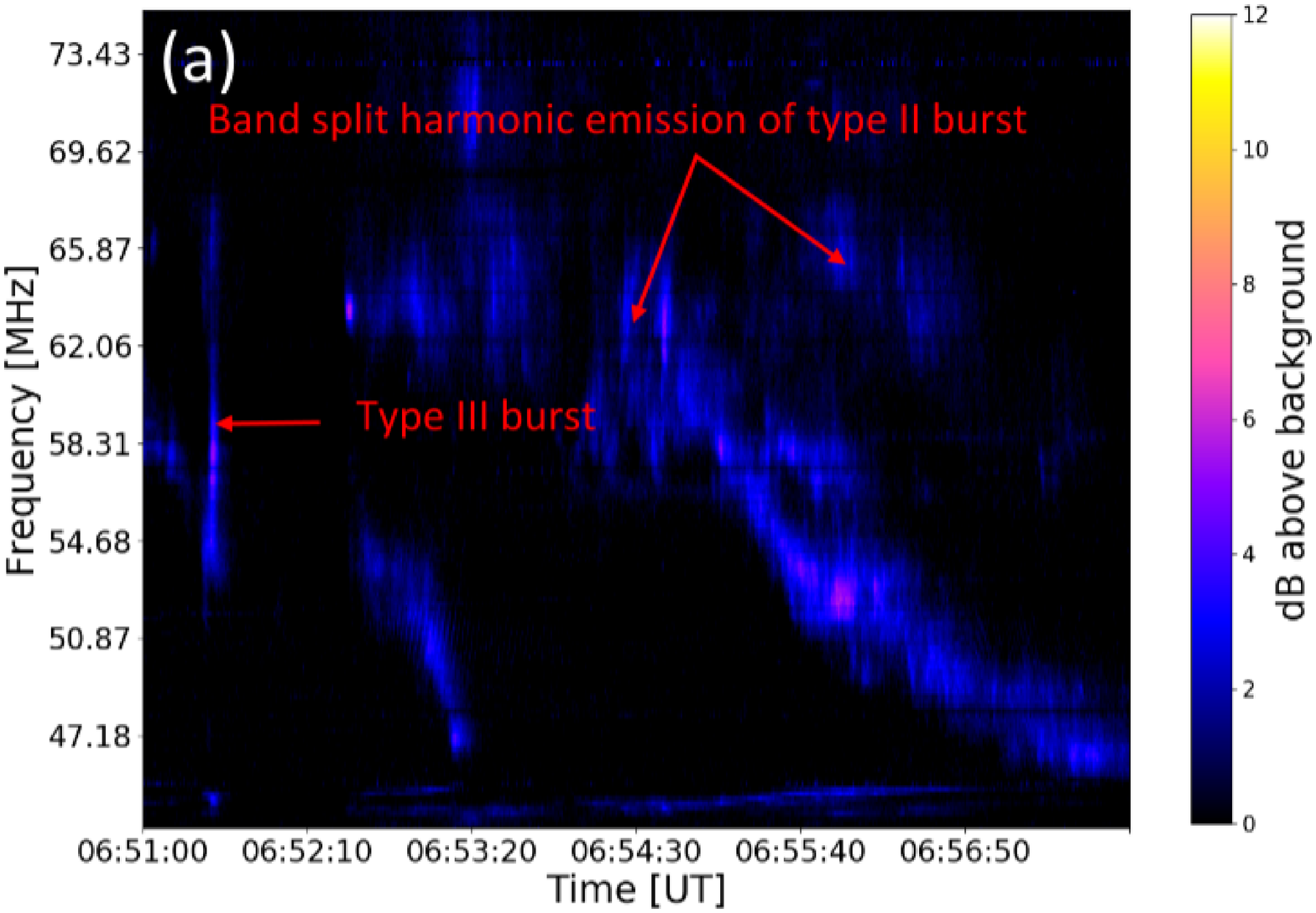}}
{\includegraphics[width=9cm,height=8cm]{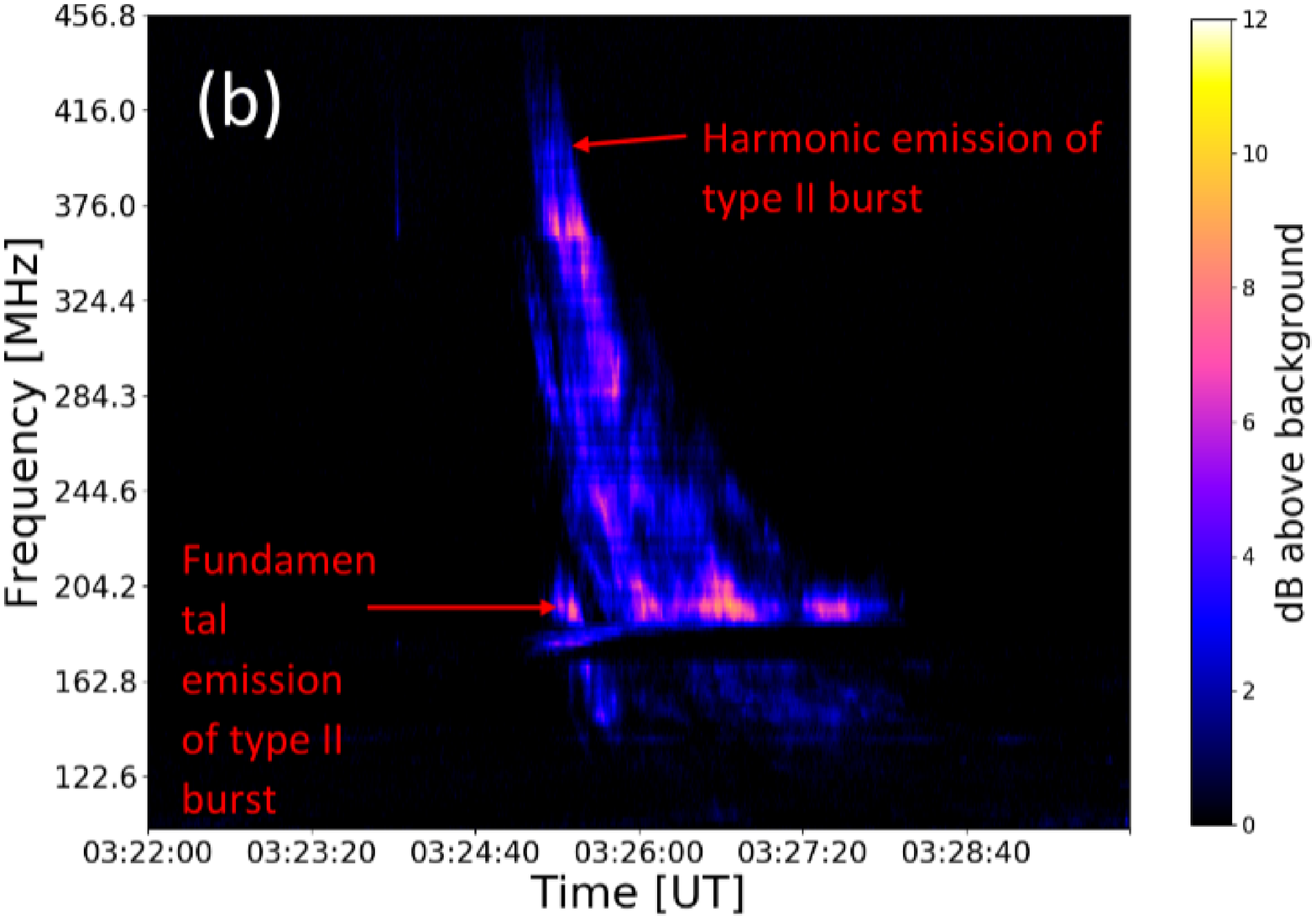}}
\end{tabular}
\caption{(a) The processed low-frequency type II burst dynamic spectrum was observed by CALLISTO RWANDA station on August 22, 2015. Its drift rate is -0.08 MHz/s. It was associated with an M1.2 flare located near the solar disk (N15E20). Its associated CME has an average shock speed of 547 km/s. (b) shows the high frequency ($ \geq $ 150 MHz ) burst observed by CALLISTO ALMATY station on November 04, 2015. Its drift is higher (-0.83 MHz/s) and consistent with its starting frequency. The associated M3.1 flare originates at N15W24 on the solar disk. The associated CME has an average speed of 272 km/s.}
\label{fig1} 
\end{figure}

\noindent
In Figure \ref{fig1}, panel (a) shows a processed dynamic spectrum of a low starting frequency burst observed by CALLISTO station in RWANDA on August 22, 2015, at $\sim$06:52-06:58  UT\footnote{\url{http://soleil.i4ds.ch/solarradio/qkl/2015/08/22/RWANDA_20150822_064503_59.fit.gz.png}}. It was processed to remove: i) the radio frequency interferences (RFIs); ii) the background using median subtraction. This is a typical type II radio burst with a band-spitting of fundamental emission \citep{Roberts1959} in the frequency range 73 MHz-44 MHz. Panel (b) shows the dynamic spectrum of an HF type II  
burst obtained from CALLISTO ALMATY station on November 04, 2015, during $\sim$03:25-03:27 UT\footnote{\url{http://soleil.i4ds.ch/solarradio/qkl/2015/11/04/ALMATY_20151104_031500_59.fit.gz.png}}. Note that harmonic lane of type II burst starts at $\sim$ 440 MHz. The detailed analysis of the event is presented in Section \ref{sect.3}.

\subsection{White-light and EUV observations}
\label{sec2.2}
\noindent
The white-light CMEs data used in this study are obtained with the LASCO onboard SOHO \citep{Brueckner1995} and the Cor1 and Cor2 coronagraphs from the Sun-Earth Connection Coronal and Heliospheric Investigation (SECCHI; \citet{Howard2008}) onboard the Solar Terrestrial Relations Observatory (STEREO). We extensively used the Coordinated Data Analysis Workshop (CDAW) CME database\footnote{\url{https://cdaw.gsfc.nasa.gov/}}, which is an online CME catalogue \citep{Yashiro2004,Yashiro2008,Gopalswamy2009cdaw}. The CMEs listed in this catalog use SOHO/LASCO white-light coronagraph data by manual detection of CMEs. We obtained information like CME linear speed within the LASCO field of view, estimated CME time, widths, and we accessed CME images and movies within STEREO. The link also guided the Geostationary Operational Environmental Satellite (GOES) to check the flare properties associated with the CMEs corresponding to the selected HF type II bursts, such as their onset time and classes. We also used observations at 171 \AA/211 \AA~ with the Atmospheric Imaging Assembly  \citep[AIA;][]{lemen2011atmospheric} onboard the Solar Dynamics Observatory \citep[SDO;][]{YASHIRO2020}. 

\section{Method}
\label{sect.3}

\subsection{Identification of type II bursts}
\noindent
Type II bursts and their quality (ranging from 1 (poor) to 5 (excellent))
were first checked from the website of the space weather prediction center (SWPC\footnote{\url{ftp://ftp.ngdc.noaa.gov/STP/swpc/_products/}}). The frequency range and observed timing were estimated from the dynamic spectra provided by CALLISTO spectrometers. Further, observationally derived parameters of HF type II bursts are used to characterize the properties of associated CMEs using the LASCO catalog. We considered the comparison of the time difference between the first appearance of the CME in the STEREO coronagraph/SDO and the observation of the fundamental band of the metric type II bursts to confirm their association \citep{Aurass}. In this study, we considered 15 minutes of difference. However, for some events, the fundamental and harmonic band could not be distinguished easily. 

\subsection{Derivation of CME speeds using SRBs physical parameters}
\noindent
The parameters obtained with radio observations, such as drift rate, can indicate CME-driven shock temporal evolution, as the type II bursts generally are associated with CME shock \citep{Gopalswamy2010a, kumari2017a}. The drift rate of type II radio burst is calculated by,

\begin{equation}
 DR [\textrm{MHz/s}] = \frac{\vartriangle f [MHz]}{\vartriangle t [s]}= \frac{f_e - f_s}{t_e - t_s}  
 \label{eq1}
\end{equation}

\noindent
where, DR, $f_s$, $f_e$, $t_s$ and $t_e$ are the drift rate, start frequency, end frequency, start time and end time of the type II burst. The frequency cut-offs and time duration are in MHz and seconds, and the drift rate is expressed in MHz/s. 
There is a relationship between the drift rate of metric type II burst, the shock speed that produces the burst, and the density gradient in the solar corona \citep{Gopalswamy2011a}.
\noindent
Therefore, the drift rate $\frac{\vartriangle f}{\vartriangle t}$ at the observed frequency can be related to the speed of the source \citep{Mann2005,aguilar2005}.
The speed of the CME/CME shock can be estimated using the expression given by  \cite{Gopalswamy2006, Gopalswamy2011a} as,

\begin{equation}
 V [km/s]= -\frac{2r}{\alpha}\frac{1}{f [MHz]}\frac{\vartriangle f [MHz]}{\vartriangle t [s]}
 \label{eq2}
\end{equation}
\noindent
where, $f$ is the type II starting frequency or emission frequency, $r$ is the shock height expressed in solar radii, Rs = 696000 km and $\alpha$ is the exponent describing density (n) variation over the radial distance (n $\sim$ $r^{-\alpha}$ ). In this study, we used  $\alpha$=7.56, which was previously derived by \cite{Gopalswamy2013} for metric type II bursts. 

\subsection{Measurement of CME shock heights at type II onset}
\noindent
We used contemporaneous data of extreme ultraviolet (EUV) solar data from the Solar Dynamic Observatory (SDO), white-light CME data from STEREO, and type II data from CALLISTO spectrographs. We did a case study on 05 October 2013 and 13 June 2010 HF type II bursts to validate the aforementioned method. We used this dataset as the CME evolution was clearly seen in the STEREO FOV.  We took proper measures to minimize the projection effect due to the location of the source region. We used the leading edge and wave diameter methods. In the wave diameter method an approximately circular EUV or white-light disturbance is observed in the images, and a circle is fitted to the outermost part of the disturbance. The radius of the circle is then taken as the CME shock height. If the shock feature in the images is not clear in the leading edge method, the CME leading front height is taken as the height at which the type II burst (shock) forms. We have taken the CME height to be the same as the shock height and the type II burst height, as in \cite{Gopalswamy2013}. 

\setlength\tabcolsep{1.5pt}

\begin{landscape}

\begin{longtable}[]{cccccccccccccc}

\caption{Details of the events:  Type II bursts, CMEs and Flares}
\label{table1}\\

\hline{SN} &   {Type II}\footnotemark[1]  & {fs}\footnotemark[2]   & {df/dt}\footnotemark[3]   & {CME}\footnotemark[4]  & $v_{l}$\footnotemark[5]   & $v_{d}$\footnotemark[6]  &$v_{II}$\footnotemark[7]& {Width}\footnotemark[8] & {r}\footnotemark[9]  & {Flare }\footnotemark[10] & {Loc}\footnotemark[11] & {Class}\footnotemark[12] & {r and $v_{II}$ }\footnotemark[13] \\
        
        &     (UT)      & (MHz)  & (MHz/s)   &  (UT)  & (km/s)    & (km/s)   & (km/s)  &   (degrees)      & (Rs) &  UT      &        &       & measurement method \\\hline
              
\endfirsthead

1 & 20100612 01:00:00- 01:03:30 &170&-0.28&01:06:04 &486 & 579&673 & 119&1.91&01:30 & N23W43 & M6.0& LE \\ 
              
2 & 20100613 05:39:00-05:45:00 & 220 &-0.25 & 05:36:08 & 320 & 245 & 522&16 & 1.17 & 05:40 & S23W75 & M1.0& LE   \\ 
   
3 & 20101103 12:15:30-12:19:00 &450 &-0.47& 12:16:08 &241 & 258&511& 66 &1.34&12:10 & S18E88 &C9.2& LE   \\ 

4 & 20110308 03:46:00- 03:51:00 & 160 &-0.33& 04:12:05& 732  & 516.5&1370& 260 & 1.36 &03:37 & S21W72  & M1.5& LE \\ 

5 & 20110607 06:38:00- 06:38:45 & 160 &-0.67 & 06:40:56 &1255 & 1557.5&986&360 & 2.02 &07:00 & S21W54 &M9.1& LE  \\ 

6 & 20110828  04:20:00 - 04:24:00  & 160&-0.25&04:24:08& 433& 420&847& 72&1.46&--&--&--& LE\\ 

7 & 20110906  01:47:00- 01:49:00 &250&-0.83& 01:50:31&782& 1027&982& 360&1.68& 01:36 & N14W07 &M5.7& LE\\               

8 & 20110906 22:19:00- 22:21:00 & 450&-1.66& 22:20:00& 575& 971& 1032 & 360&1.43&22:45 & N14W16 &X8.2& LE\\  

9 & 20111001  09:07:00-09:11:00&170&-0.29& 09:15:30 &448& 575&974 & 203 &1.83 &09:45& N10W06 &M1.0& LE\\ 

10 & 20111114  09:24:00- 09:26:00 &170&-0.41&09:25:30 &553 & 666&880 & 10 &1.50&09:50& N20W51 &C9.0& LE\\ 

11 & 20120517 01:33:00 - 01:34:30 &170&-0.88&01:48:05& 1582& 1811&2220 & 360 &1.90&01:25& N13W87 &M5.1& LE\\ 

12 & 20120603 17:50:00 - 17:55:00&195&-0.28& 17:55:21&  605 &367.5&793&180&1.39&17:55& N16E38 &M3.3& LE\\ 

13 & 20120607 20:03:00 - 20:09:00 &180&-0.22&20:05:30  &  494&439&639&173&1.95&20:06& S19W05 &M2.1& LE\\ 

14 & 20120608 03:08:00-03:15:00& 211 &-0.18& 03:10:30& 316 & 253 &673&  94& 1.61 & 03:20& S09W21 &M1.0& LE\\ 

15 &20120702 05:10:00- 05:13:30 & 170&-0.19&  06:00:04& 251& 276 &626& 90&1.34&06:00&N17E02&C8.0& LE\\

16 & 20120728  20:52:00 - 21:00:00 &400&-0.41& 20:55:30 & 420 & 287 &504& 360 & 1.52 & 20:56 & S25E54 & M6.1& LE\\  

17 &20120806   08:20:30 - 08:24:00 &160&-0.14& 08:36:06&242& 285&520& 56&1.77 &08:35& S12E59 &C5.9& LE\\ 

18 & 20121121  07:01:00- 07:09:00&360&-0.250& 07:06:00&297&210&390&70&1.51 &06:56& N06E10 &M1.4& LE\\    

19 &20130113   08:40:00- 08:42:00 & 300&-1.08&08:48:05& 696& 829 &929& 46&1.25&09:10& N17W22 &M9.1& LE\\

20 &20130411   07:03:00 - 07:05:00 &160&-0.50&07:10:00 &861& 834&1096& 360&1.45 & 06:55 & N10W01 &M6.5& LE\\ 

21 & 20130502  05:07:00-05:13:30 &170&-0.33 &05:24:05& 671 &608&1186&99&1.70&05:30& N10W28 &M1.0& WD\\ 

22 & 20130517  08:52:00-08:55:00 & 270&-0.97 &08:50:27& 1345 &992 &1183& 360& 1.50 & 08:43 & N12E22 & M3.2& LE\\

23 &20131005   06:58:30- 07:00:00  &170&-0.66& 06:55:30 &964& 1229.5&1450& 360&1.72&--&--&--& LE\\ 

24 &20131011   07:11:00- 07:14:00& 250& -0.94&07:12:07& 1200 & 928&964 &360&1.34&07:10& S08E20 &C9.0& LE  \\ 

25 &20131025   08:00:30 - 08:09:00 & 240& -0.37&08:12:05& 587 & 451 &893& 360&1.59&08:10& N06W24 &C6.0& LE\\  

26 &20131107   03:40:00- 03:42:00 &250&-0.41&04:05:31 &373& 486 &526&69 &1.61 &03:34& S11E10 &M2.3& WD\\ 

27 &20131107   14:25:00-14:27:00 &260&-0.50& 14:45:00&411& 443 &626 & 360 & 1.25 &15:17& S12E21 & M2.4& LE\\ 

28 &20131207   07:26:00 - 07:30:00 & 240&-0.5&07:36:05& 1085 & 522 &1276&360&1.36&07:35& S13W11 &C2.0 & LE\\ 

29 &20131212   03:19:00- 03:22:00 & 150&-0.45&03:20:08 & 1002 & 806.5 & 1100&276 & 1.46 & 03:20& S23W46 &C3.0& WD  \\

30 &20140106   07:47:00 - 07:49:30 & 160&-0.60&08:00:05& 1402& 1036 & 1579&360&1.50&07:30& S13W83 &C2.1& LE\\ 

31 &20140108   03:48:30-03:52:00 & 240&-0.57&03:48:09& 643& 560 &535&108& 1.28&03:39& N11W91 &M3.6& LE\\ 

32 &20140126   08:36:30- 08:38:00 &150& -0.56& 08:36:05&1088& 1251 &1368&255& 1.82&09:00& S17E89 &C3.1& LE\\ 

33 &20140211   13:27:00 - 13:30:00 &450 &-1.10&13:48:05& 330 & 711 & 851& 208& 1.58 &13:50& S11E11 &C9.0& LE\\ 

34 &20140220   07:49:45-07:49:00& 300&-1.55 &07:45:57& 948& 1122.5&995&360& 1.18 & 07:26 & S14W81 & M3.0& LE\\ 

35&20140306    09:27:00- 09:28:30 &220&-0.55& 09:30:53&252& 635 &476& 56&1.38&06:30& N15W36 &C3.0& LE\\ 

36 &20140822   10:20:00- 10:21:30  &170&-0.50&11:12:05&217&-- &--& 38&--&11:10& N10W07 &C2.2&--\\ 

37 &20140928   02:47:00-02:56:00  & 200&-0.18& 02:48:09&215& 232& 493&60&1.40& 02:48 & S14E24 & M8.0&WD\\
 
38 &20141030    13:08:00-13:13:00  & 200& -0.23&13:08:00& 285 &330 & 541&112& 1.56 &12:37& S12W91 &C2.9& LE\\ 

39 &20141102    09:45:00-09:47:00  & 160&-0.33&10:00:06& 461& 605 & 555 &$\succeq168$&1.45&09:20& S12W91 &C4.5& LE\\

40 &20150311    16:24:00-16:30:00 & 200&-0.19&17:00:05&240& 224 &638&74& 1.28&16:11& S16E13 &X2.1&WD\\ 

41& 20150601    13:32:00-13:34:30 &200&-0.53&13:36:00&748& 654& 715& 360& 1.34&--&--&--& LE\\ 

42 &20150822    06:50:00-06:52:30 &155&-0.30&07:12:00&547& 641.5& 601& 360&1.80&06:39& S14E09 &M1.2& LE\\

43 &20151017    04:22:00- 04:26:00  & 165&-0.18&04:36:00&218& 325& 561& 70&1.62&05:20& S10E33 &C6.5& LE\\ 

44 &20151104    03:24:00- 03:29:00 &450 &-0.83 &04:00:04& 272&-- &-- &64&--&04:00& N15W24 &M3.1&--\\ 

45 &20151104    12:03:00-12:08:00 & 410&-0.70&12:36:04&283& -- &-- &64 &--&12:30& N05W04 &M8.0&--\\ 

46 &20160710    01:00:00-01:03:00  & 260&-0.44& 01:05:30&368& 461 &534& 101&1.48&00:53& N09E49 &C8.6& LE\\ 

47 &20170403    14:25:30- 14:28:30 &350& -0.8&14:25:30&471& 518&  580 &80& 1.23&14:24:00& N15W76 &M8.5& LE  \\ 

48 & 20171018   05:38:00- 05:45:00 & 200    & -0.53 & 05:40:00 & 1576   & 688 &580  & 360 &  1.41   &--&--&--& WD\\ 

49 &20190421    04:46:00- 04:56:00 & 152    & -0.26 &  04:48:00 & 463  & 794    & 565 &  274     & 1.43 &--&--&--& LE\\ 

50 &20190422    02:52:00- 03:00:00 & 162    & -0.34     & 03:12:00  & 422  & 981.5  & 567&  269 & 1.21  &--&--&--& LE\\ 

51 &20190506    05:12:00- 05:15:00 & 165   & -0.51    & 05:20:30 & 239  &1053 &950 &  53       & 1.85    &  05:18 & N09E47  & M1.0& LE \\\hline  

\footnotetext[2]{$f_{s}$ = Highest start frequency of type II bursts, which can be either fundamental and/or harmonic}
\footnotetext[3]{df/dt = Drift rate} 
\footnotetext[5]{$v_{l}$ = Average CME speed within LASCO FOV} 
\footnotetext[6]{$v_{d}$ = CME  shock speed derived using parameters from dynamic spectra} 
\footnotetext[7]{$v_{II}$ = CME shock speed at type II onset}
\footnotetext[9]{r = CME shock heights at type II onset} 
\footnotetext[11]{Loc = Solar flare heliographic coordinates} 
\footnotetext[1]{UT = Universal time} 
\footnotetext[9]{$R_{s}$ = Solar radii} 
\footnotetext[13]{LE = Leading Edge}
\footnotetext[13]{WD = Wave Diameter}

\end{longtable}
\end{landscape}

\section{Results and discussions}
\label{sect.4}
\noindent
We prepared a list of HF type II bursts from SWPC event lists and their associated CMEs and flares from 2010-2019. Table \ref{table1} contains the list of these events. Columns (2 to 4) list the type II parameters such as type II onset and end times (duration), type II upper-frequency cut-off, and the drift rate of type II bursts. Columns (5 to 10) lists the associated CME characteristics such as the CME shock time, its average linear speed within LASCO FOV ($v_{l}$), CME average speed derived using parameters estimated from the CALLISTO dynamic spectra ($v_{d}$), the average CME shock speed at type II onset ($v_{II}$), CME width and CME heights at type II onset (r).  Columns (11 to 13) list the flare onset time, flare location, and the flare class, respectively. The last column shows the measurement methods used to find the shock heights and CME speeds at type II onset.

\subsection{Occurrence of high frequency type II bursts}
\begin{figure}    
   \centering\includegraphics[width=0.49\textwidth,clip=]{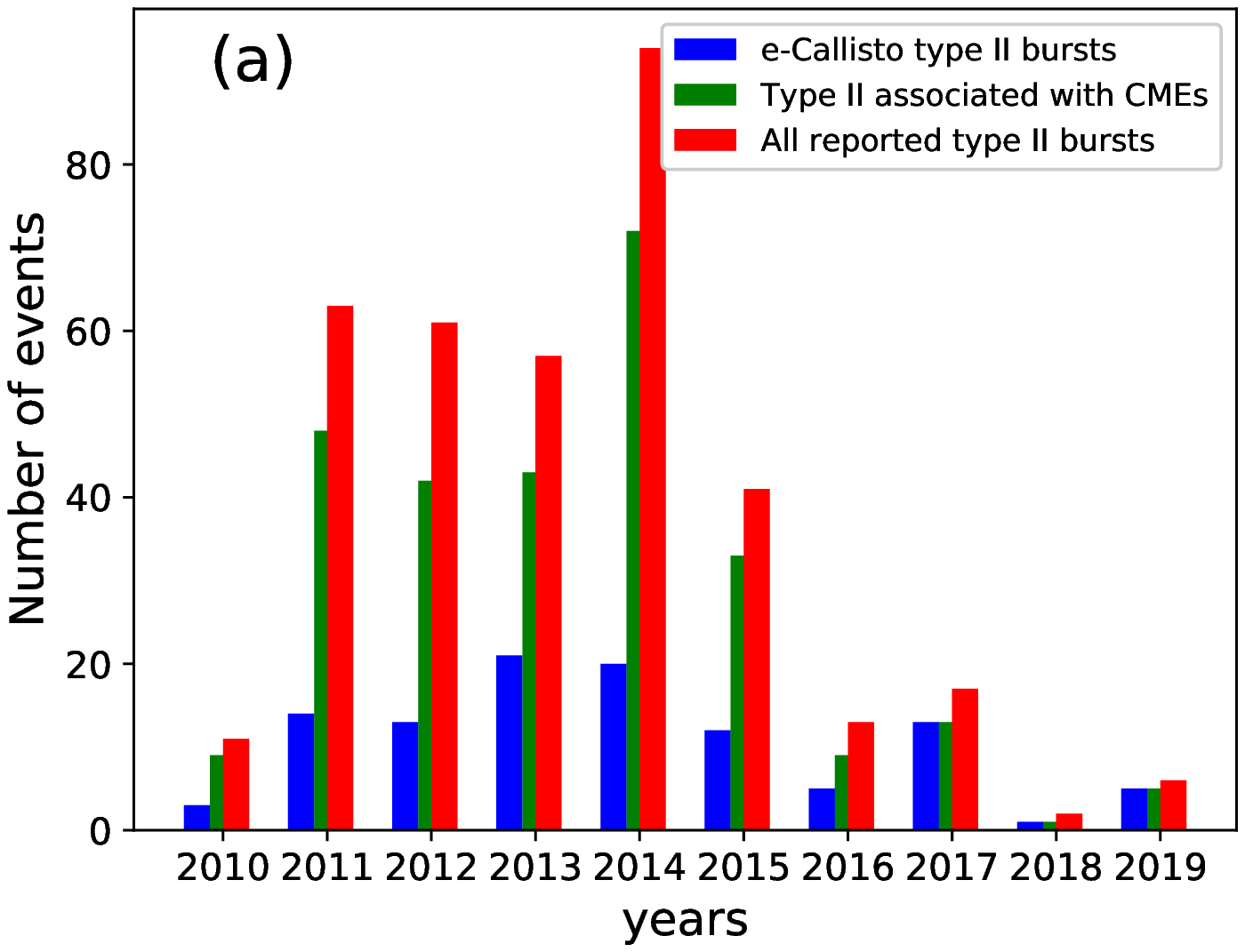}
   \centering\includegraphics[width=0.49\textwidth,clip=]{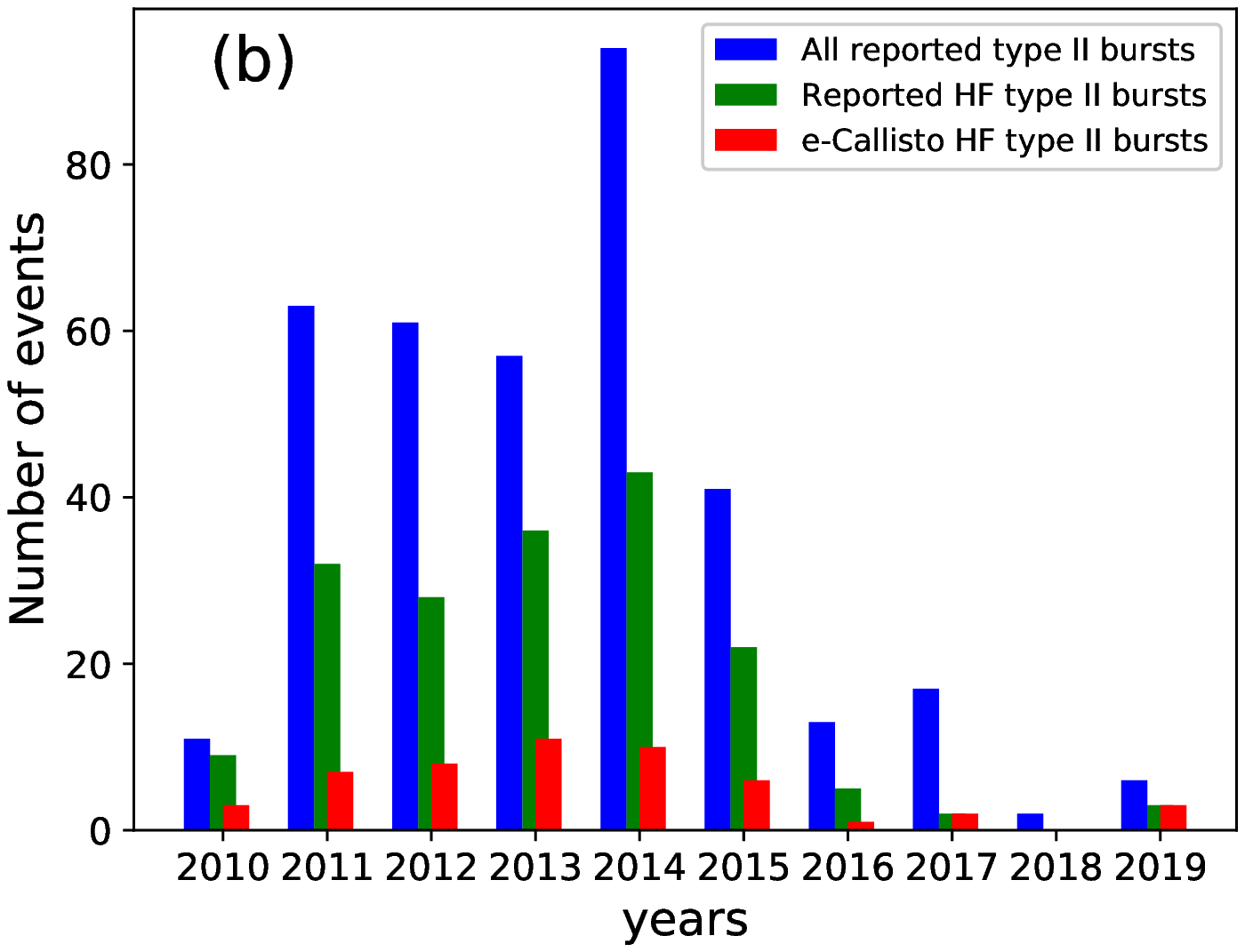}
   \centering\includegraphics[width=0.49\textwidth,clip=]{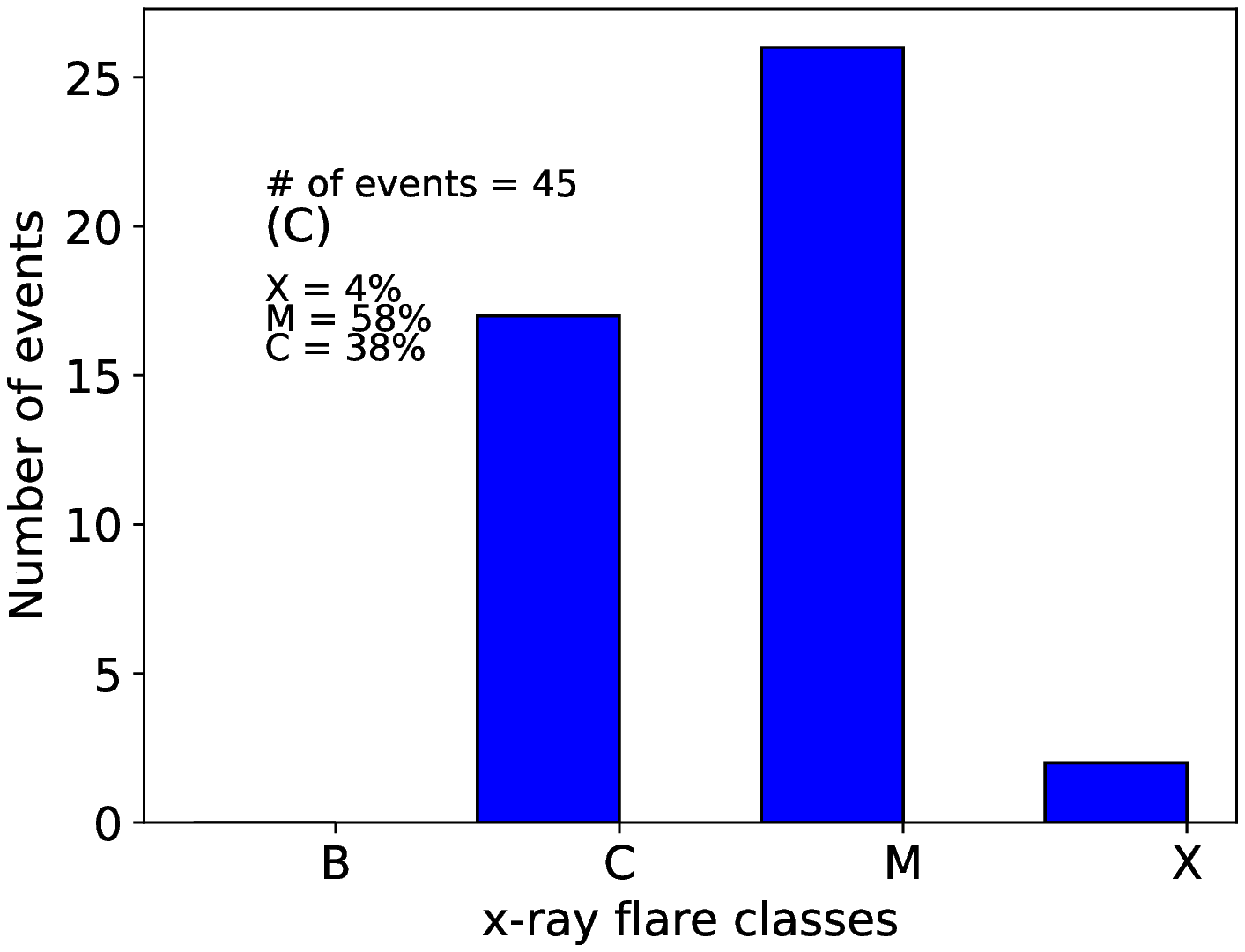}
   \centering\includegraphics[ width=0.49\textwidth,clip=]{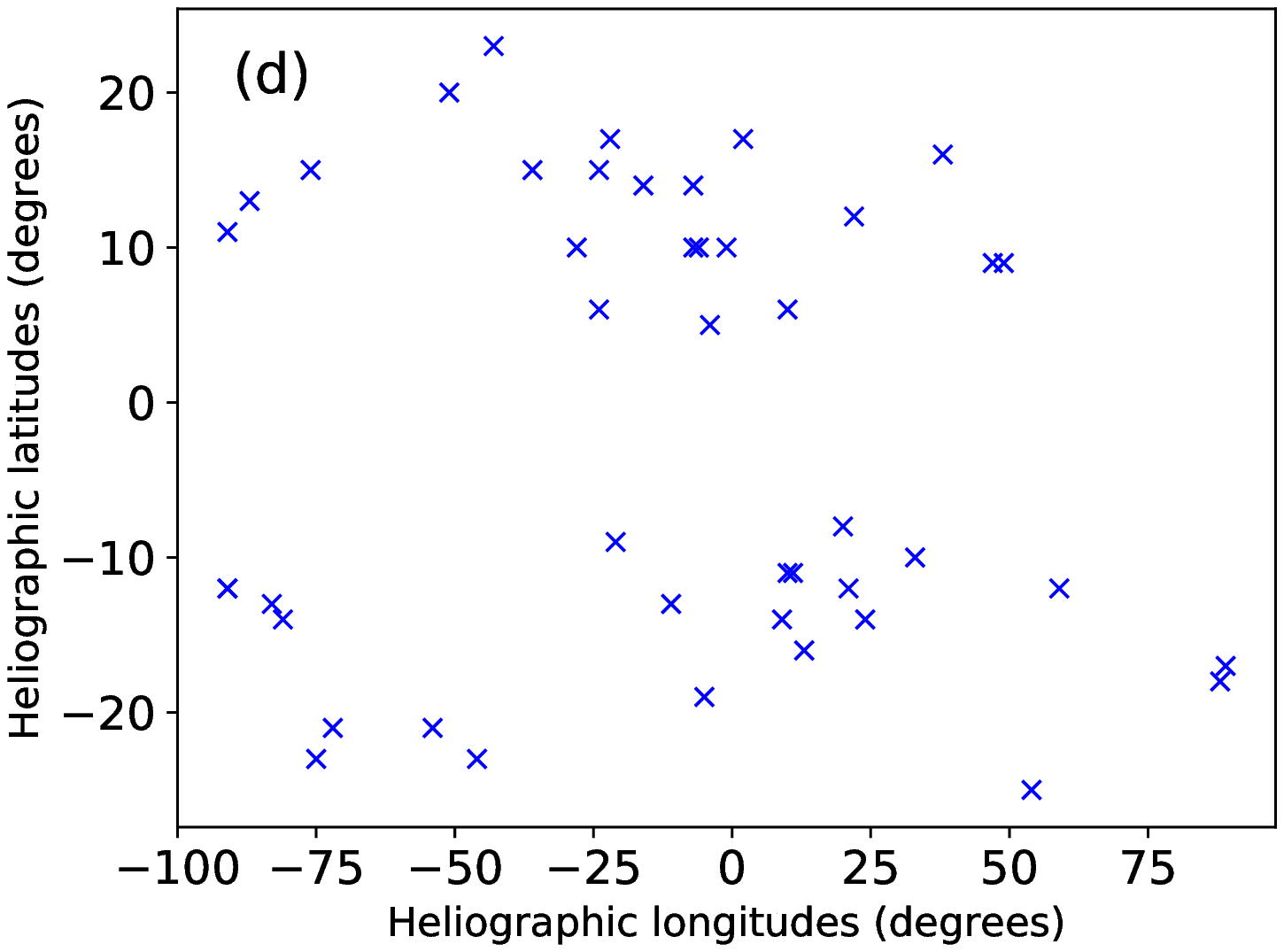}
              \caption{(a) Distribution of 
              type II bursts, type IIs associated with the CMEs and type II bursts observed with e-Callisto network per year from 2010-2019 represented by  `red', `green' and `blue' colors, respectively.
              (b) type II bursts, HF type IIs and HF type II bursts observed with e-Callisto network per year from 2010-2019 represented by  `blue', `green' and `red' colors, respectively;
              (c) X-ray flares class associated with HF type II radio bursts; and 
              (d) Heliographic longitude Vs heliographic latitude of flares associated with the high-frequency type II bursts.
                      }
\label{fig2}
\end{figure}

\noindent
From 2010 until 2019, the SWPC event list reported a total of 365 type II bursts. Out of these type II bursts, only 107 type II bursts were detected with an e-Callisto spectrometer. Among these 365 type II bursts, 84 ($\sim 23\%$) are not associated with CMEs. Figure \ref{fig2}a shows the distribution of occurrence of type II bursts, type IIs associated with the CMEs, and type II bursts observed with the e-Callisto network per year from 2010-2019. Figure \ref{fig2}b shows the distribution of occurrence of all type II bursts, HF type II bursts, and HF type II bursts observed with the e-Callisto network per year in the same period. There is a total of 180 HF type II bursts reported. We selected 51 ($\sim 28\%$) HF type II bursts observed with e-Callisto solar spectrometers. Some CALLISTO instruments cannot track the Sun, reducing potential observation time to about 4 hours per day. The CALLISTO in Rwanda falls in this category.
Moreover, e-Callisto, in principle, can only detect the radio bursts which radiate in the design frequency range of CALLISTO, ranging between 45 MHz and 870 MHz, i.e., meter wavelength radio bursts.  Also, It cannot see metric bursts behind the Sun, as they are formed at low heights. If the bursts continue to higher heights, they can be detected by space-borne instruments (deca-hectometric or kilometric bursts, also known as DH and km type II bursts). Results from this study indicate that all 51 HF type II bursts analyzed are associated with CMEs in agreement with previous studies by \cite{Gopl2,Ramesh,Cho} who found that the formation of type II bursts in the low corona are explained by the CME-driven shock, the analysis considered different methods. Most of the type II bursts were reported in 2014, which was also the peak of solar cycle 24. Figure \ref{fig2}c shows the distribution of X-ray flare class associated with the HF type II bursts observed with e-Callisto. Out of these shortlisted 51 radio bursts,  45 were accompanied by X-ray flares. Most of the HF type II bursts are associated with M ($\sim 58 \%$) and C ($\sim 38 \%$) class X-ray flares, respectively. Only ($\sim 4\%$) HF type II bursts are associated with X class flares, and none of these bursts are accompanied by B class flares. Figure \ref{fig2}d shows the heliographic longitudes and latitudes of the flares associated with the type II bursts and it is found that all of them originate close to the equator (heliographic latitudes $\pm$ $25^{\circ}$), the similar results were obtained by  \cite{Mahender2020} who analyzed 426 type III bursts and found that most of them originate in heliographic coordinates $\pm 23^{\circ}$.\\

\noindent Figure \ref{figLOLA}a shows the variation of the starting frequency with the heliographic latitudes. Blue and red colors are indicating the northern and southern latitudes, respectively. Figure \ref{figLOLA}b illustrates the variation of the analyzed high-frequency type II bursts with the heliographic coordinates of the associated flares. The blue and red colors stand for the western and eastern longitudes. Due to the magnetic field connection governed by the Parker spirals and directivity of type II bursts \citep{Gop2016, Sas2013, Mahender2020}, presumably, we might have observed more number of type II bursts originated from the western longitude (27/45), with a high probability to produce a Solar Energetic Particle (SEP) event. The points above the dashed line in Figures \label{figLOLA} a and b indicate the variation of the higher-starting-frequency ($\geqq$ 300 MHz) bursts with the heliographic latitudes and longitudes, respectively.  Similarly, 6/10 (60\%) of the bursts originate from the western heliographic longitudes and range between $-25^{\circ}$ and $+25^{\circ}$ at the latitudes. In general, the analyzed high-frequency type II bursts are close to the disk center, similar to the previous study by \cite{Singh} on the analysis of type III using the statistical method.   

\begin{figure}    
   \centering\includegraphics[width=0.49\textwidth, clip=]{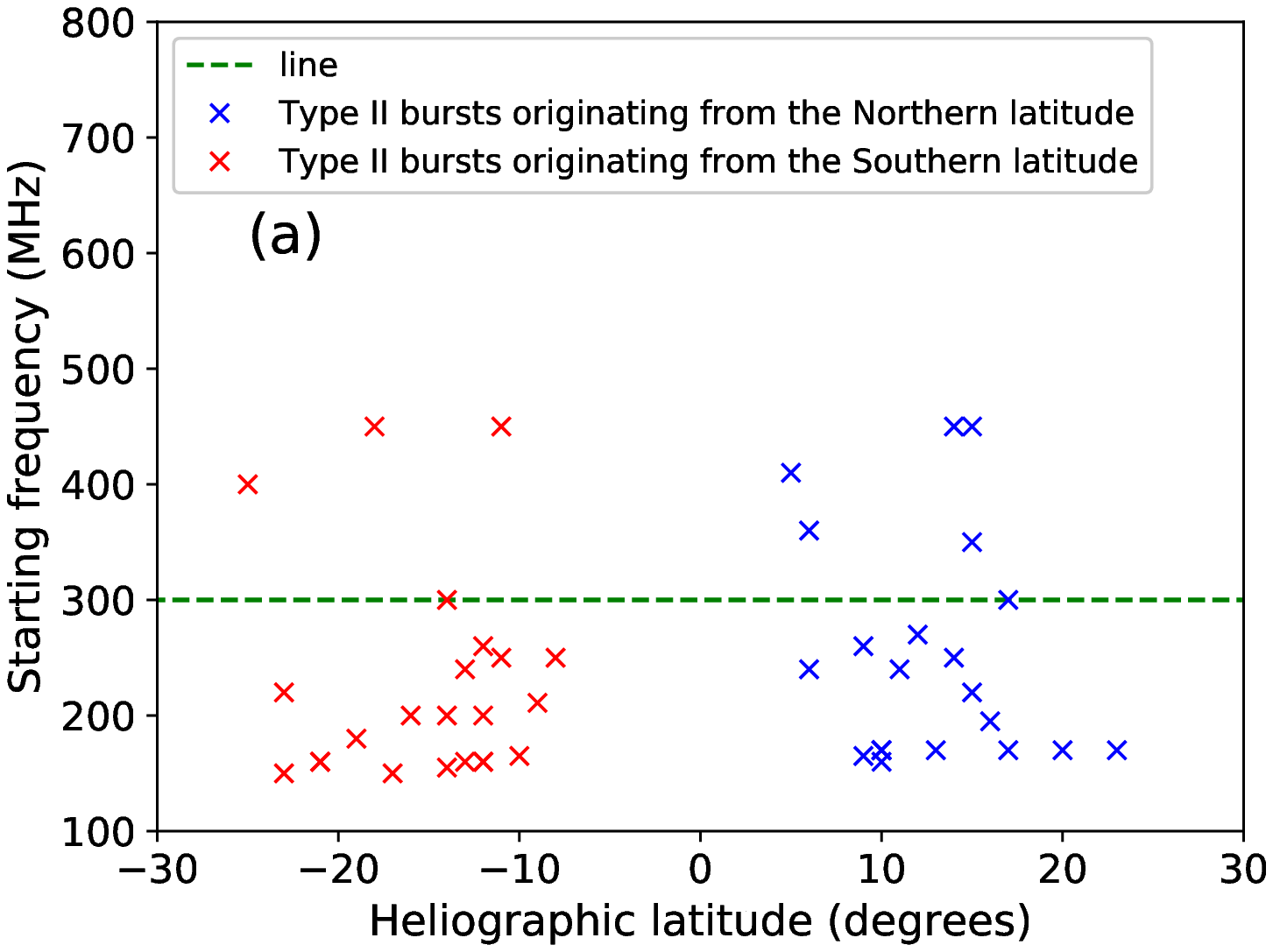}
   \centering\includegraphics[width=0.49\textwidth, clip=]{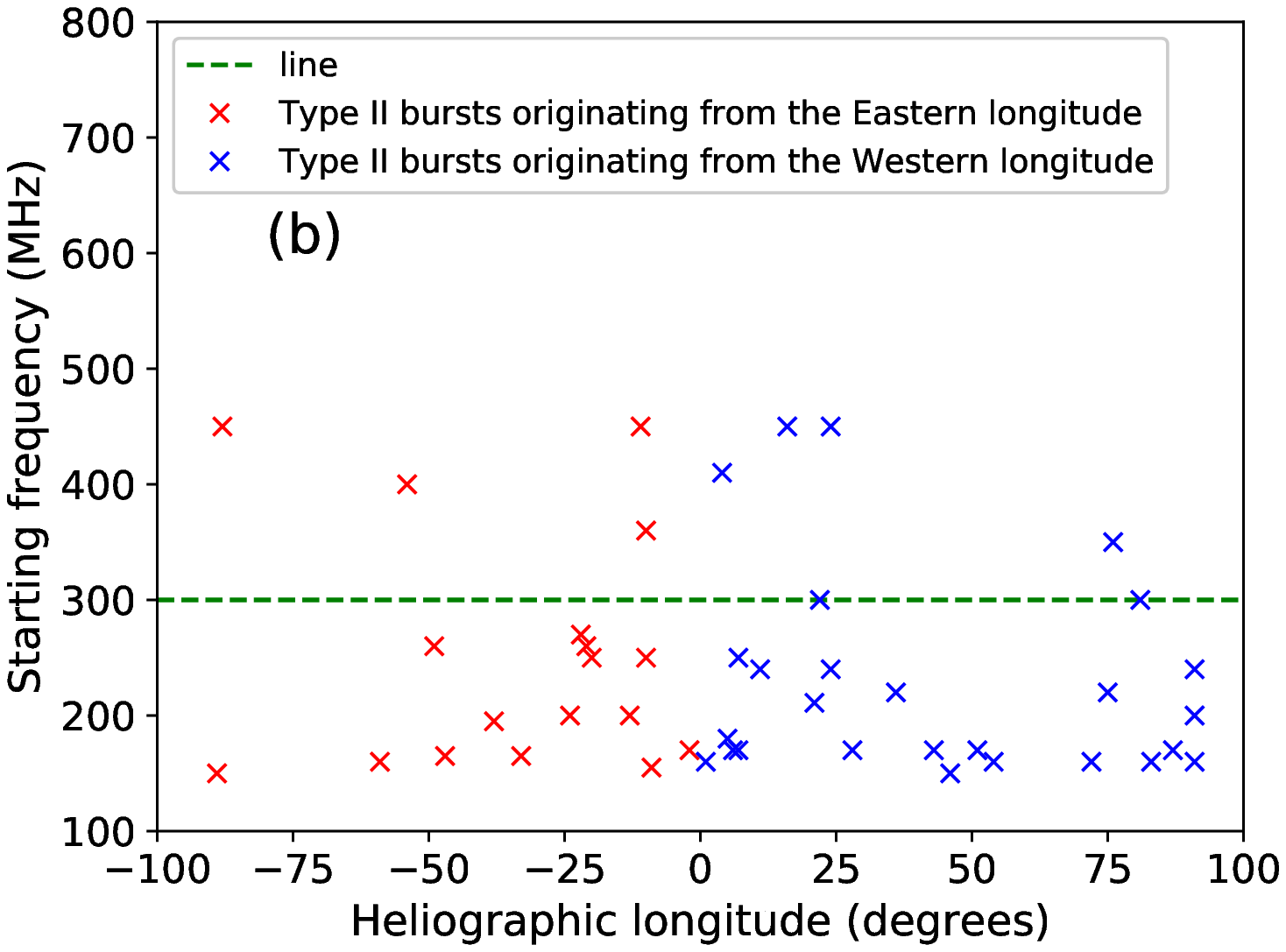}
\caption{(a) The variation of the analyzed high frequency type II bursts with the heliographic latitude. (b) The variation of the analyzed high frequency type II bursts with the heliographic longitude.  The green dashed lines on both Figures (3 a and b) indicate the selected events with higher-starting frequency ($\geq$ 300 MHz).}
\label{figLOLA}
\end{figure} 

\subsection{SEPs and the analyzed high starting frequency type II bursts}
\noindent
The same shock produces the type II radio bursts and the accelerated ions; thus, one would expect the association between type II bursts and SEP events \citep{Kahler, Cane1990}. In this study, nine events out of 51 are accompanied by SEP events. Table \ref{table2} lists the 9 CMEs associated with the analyzed high starting frequency type II bursts accompanied by SEP enhancements. The table contains the following information: Type II date, starting frequency (MHz), the average speed of the associated CME obtained from the LASCO catalog list, CME speeds derived using type II parameters estimated from the dynamic spectra, CME speeds at type II onset, CME angular width, heliographic coordinates of the associated flare, the associated flare class, SEP intensity (Ip in pfu), and the approximate duration of the SEP event in days. The analysis found that 8/9 CME events accompanied by SEP events are fast, wide, and originate from the western hemisphere. Of the 9 events, 7 are associated with M class flare. The average speed at type II onset (1118 km/s) of the SEP- associated CMEs is around two times the average speed at type II onset of all analyzed CMEs. Among the 9 SEP-accompanied CMEs analyzed in this study, 5 are halo CMEs, 3 have widths greater than $100^{\circ}$ except the 20131107 events originating from the Eastern hemisphere, which is having a width of $69^{\circ}$.  The present study confirms the previous results \citep{Gopalswamy2002, Gopalswamy2003}. The Proton, Height-Time, X-ray (PHTX) plots of the SEP-accompanied events can be accessed on \url{https://cdaw.gsfc.nasa.gov/CME_list/daily_plots/sephtx/} and more information about the PHTX plots can be found in \cite{Gopalswamy2009}. 

\setlength\tabcolsep{1.5pt}

\begin{landscape}

\begin{longtable}[H]{ccccccccccc}

\caption{Type II bursts associated with SEP events}
\label{table2}\\

\hline{SN} & Type II & fs & $v_{l}$ & $v_{d}$ & $v_{II}$ & Width & Loc & Flare & $Ip^{a}$ & Dur $(days)^{b}$ \\
           &  Date   & MHz & km/s    & km/s   & km/s     & deg   &     & Class &     &     \\\hline
              
\endfirsthead 

1 & 20110308 & 160 & 732  & 516.5&1370& 260  & S21W72  & M1.5 & 60& F \\ 

2 & 20120517 &170& 1582& 1811&2220 & 360 & N13W87 &M5.1 & 120 & F \\ 

3 &20130411   &160&861& 834&1096& 360& N10W01 &M6.5 & 100 & F\\

4 & 20130517   & 270& 1345 &992 &1183& 360 & N12E22 & M3.2 & 13 & HiB\\

5 &20131107    &250 &373& 486 &526&69 & S11E10 &M2.3 & 8 & 2\\ 

6 &20140106    & 160& 1402& 1036 & 1579&360& S13W83 &C2.1 & 60 & F\\ 

7 &20140108    & 240  & 643& 560 &535 &108 & N11W91 &M3.6 & $>$ 1000 & HiB and F \\ 

8 &20140220  & 300& 948& 1122.5&995&360& S14W81 & M3.0 & 12 & 2\\

9 &20141102     & 160& 461& 605 & 555 &$\succeq168$& S12W91 &C4.5 & 10 & F\\\hline

\footnotetext[1]{$^{a}$ Proton intensity in pfu (particle flux unit); 1 pfu = 1 particle per ($cm^{2}$ s sr). $^{b}$ Approximate duration of the SEP events in days. F denotes events followed by other events, and HiB denotes events occurring during the elevated SEP background due to preceding events. For F and HiB, it is difficult to estimate the number of days of the  SEP event.} 
\end{longtable}
\end{landscape}
 
\subsection{CME height and high-frequency type II burst onset}
\begin{figure}    
\centering\includegraphics[width=0.49\textwidth,clip=]{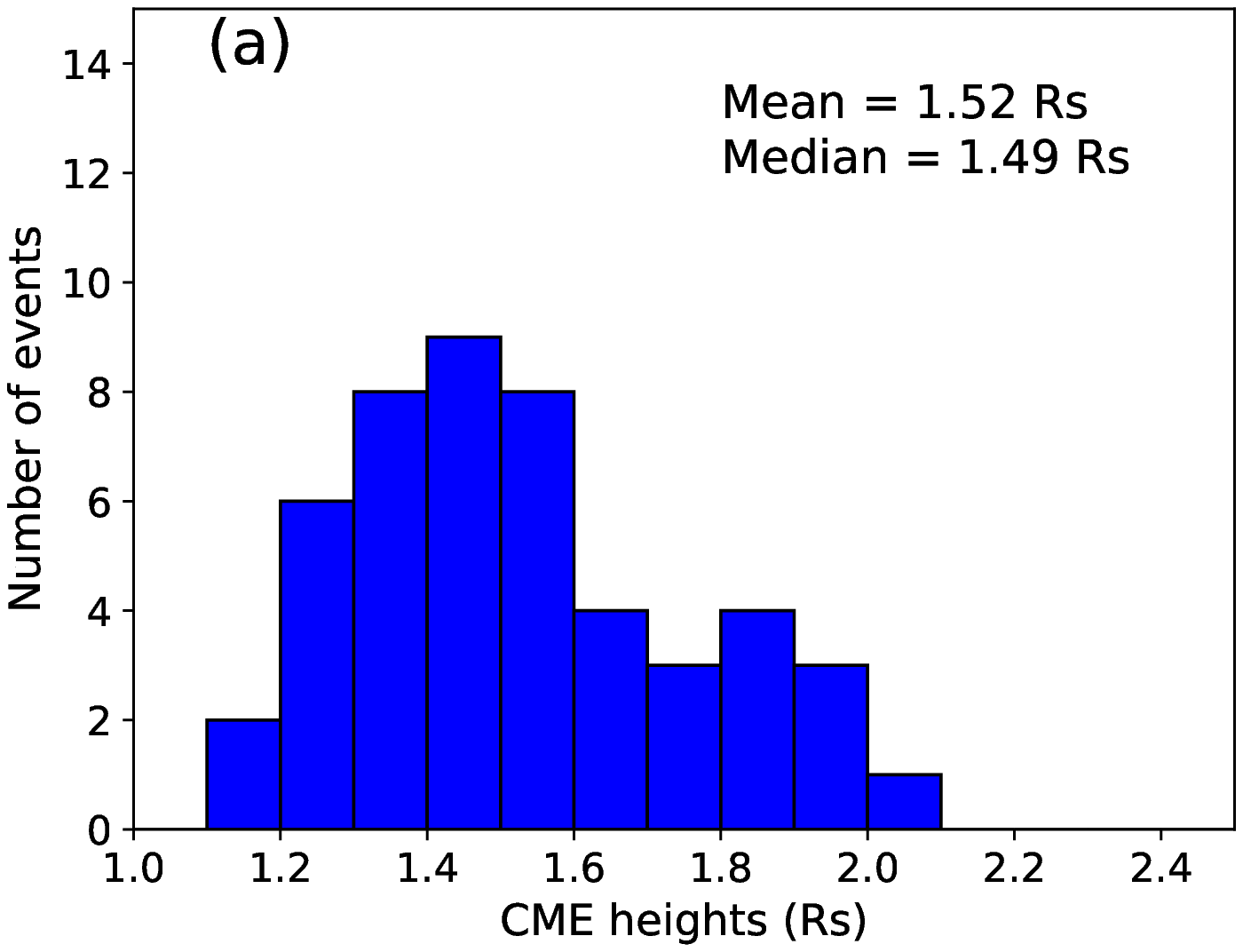}
\centering\includegraphics[width=0.49\textwidth,clip=]{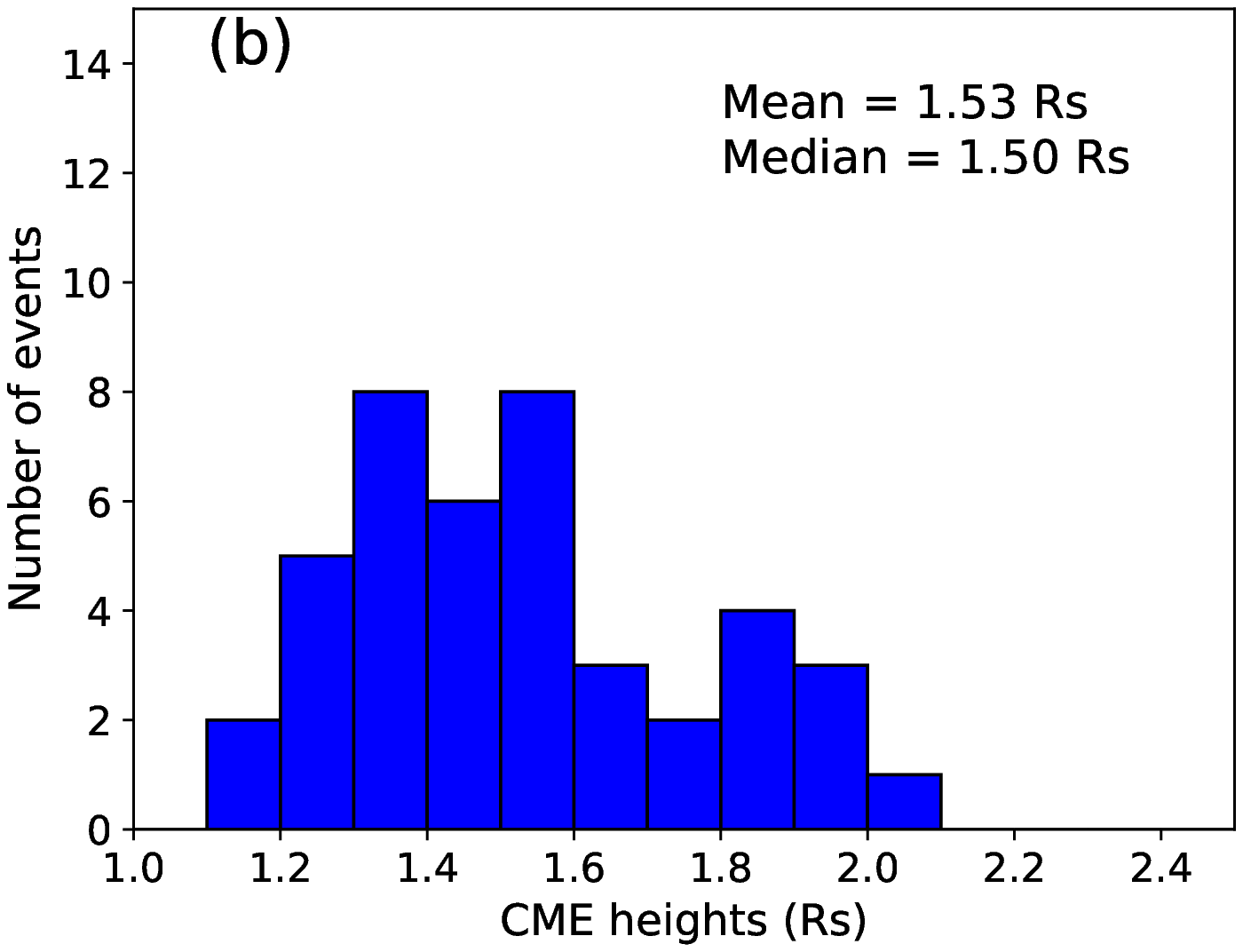}
\centering\includegraphics[width=0.49\textwidth,clip=]{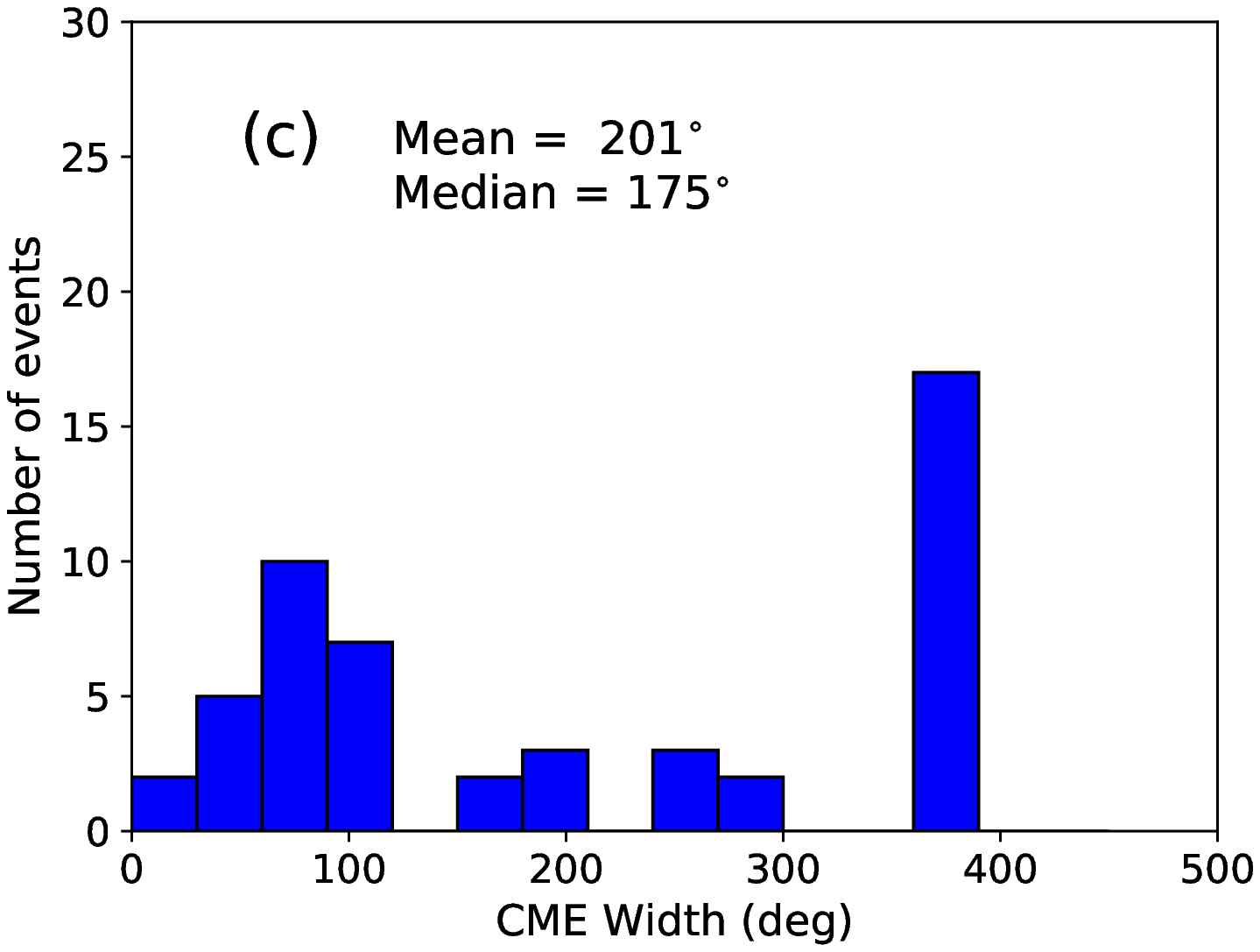}
\centering\includegraphics[width=0.49\textwidth,clip=]{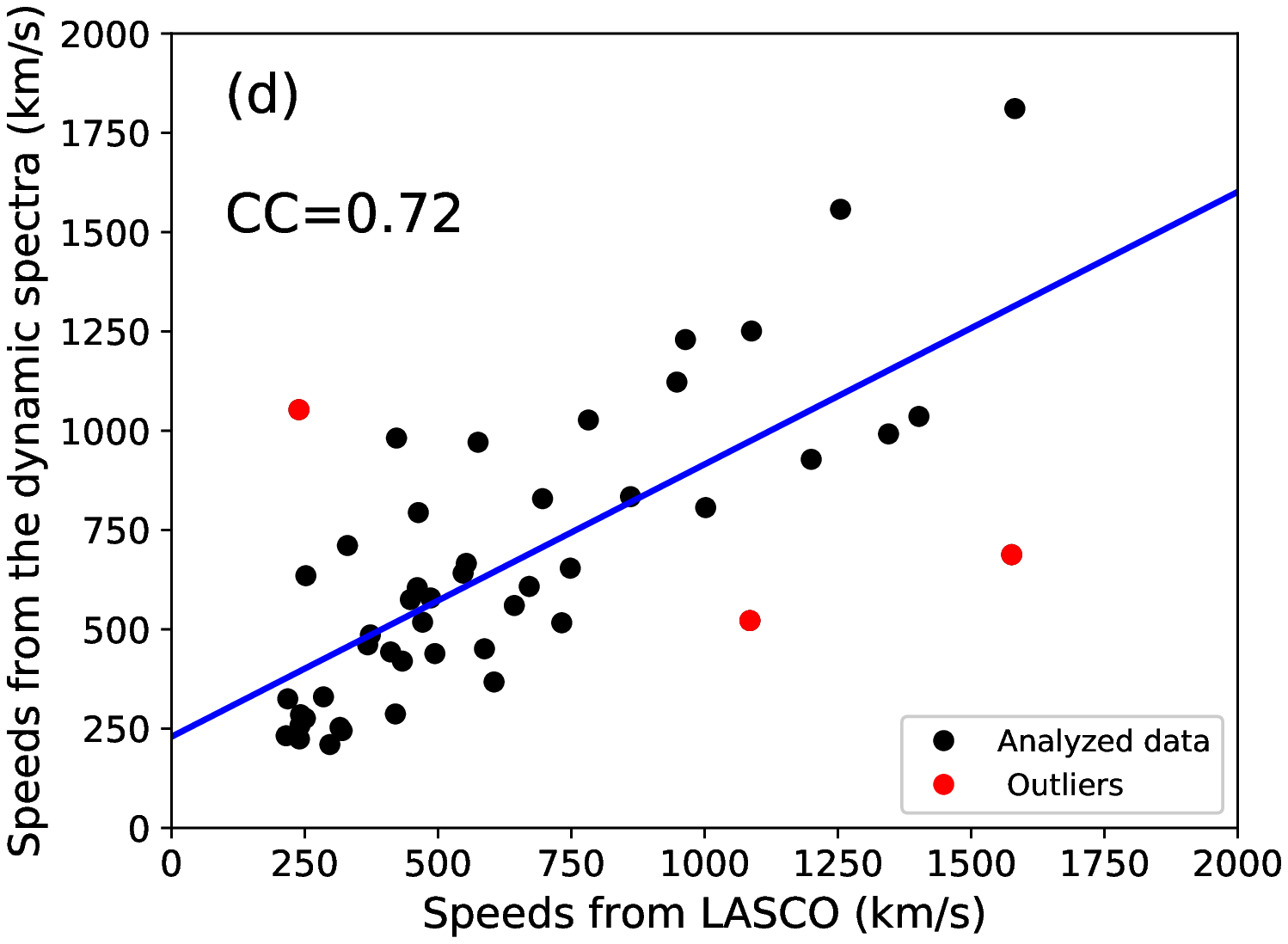}
\caption{(a) The distribution of CME heights for all analyzed events, the mean and median are 1.52  and 1.49 Rs, respectively. (b) The distribution of CME heights at type II onset identified using the leading edge method. The mean and median are 1.53  and 1.50 Rs. (c) The distribution of CME widths associated with the analyzed type II bursts. (d) The correlation between the average CME speed within LASCO ($v_{l}$) and the speed estimated using corona density model as shown in equation \ref{eq2} ($v_{d}$) for the 48 events for which the shock could be identified. The correlation coefficient (cc) is $\sim 0.72$. By excluding the three outliers (in red), the cc improves to $\sim 0.84$. } 
\label{figHeight}
\end{figure}

\noindent
Type II burst starting frequency indicates the distance from the eruption center where the electrons start to accelerate. The high starting frequency corresponds to the shock formation closer to the Sun. Figure \ref{figHeight} (a) is the distribution of CME heights for all analyzed events. The mean and median are 1.50  and 1.49 Rs, respectively. Figure \ref{figHeight} (b) shows the distribution of the CME heights for the 42 limb events (measured using the leading edge method to avoid projection effects). The mean and median CME heliocentric distances are 1.53  and 1.50 Rs, respectively, at the onset of HF type II bursts.  The CME heights obtained in this study looks smaller compared to the heights obtained in \cite{Gopalswamy2006} and comparable to the CME heights obtained by \cite{Umuhire2021} who analyzed 40 high starting frequency type II bursts and found the mean and median CME heights of 1.44  and 1.34 Rs. This study confirms the physical relationship between type II bursts and CMEs due to the similarity between type II bursts heights and associated CME leading edge heights. Figure \ref{figHeight} (c) is the distribution of CME widths associated with the high-frequency type II bursts analyzed in this study. It was found that the CMEs associated with the analyzed bursts are wide and energetic with the mean and median of $201^{\circ}$ and $175^{\circ}$ \citep{Gopalswamy2008}.  Figure \ref{figHeight} (d) is the correlation between the average CME speed within LASCO field of view ($v_{l}$) and the speed estimated using corona density model as shown in equation \ref{eq2} ($v_{d}$) for the 48 events for which the shock was identified. The correlation coefficient (cc) is $\sim 0.72$. By excluding the three outliers (in red), the cc improves to $\sim 0.84$ 

\subsection{Estimation of CME speed}
\begin{figure}
\centering \includegraphics[width=0.49\textwidth]{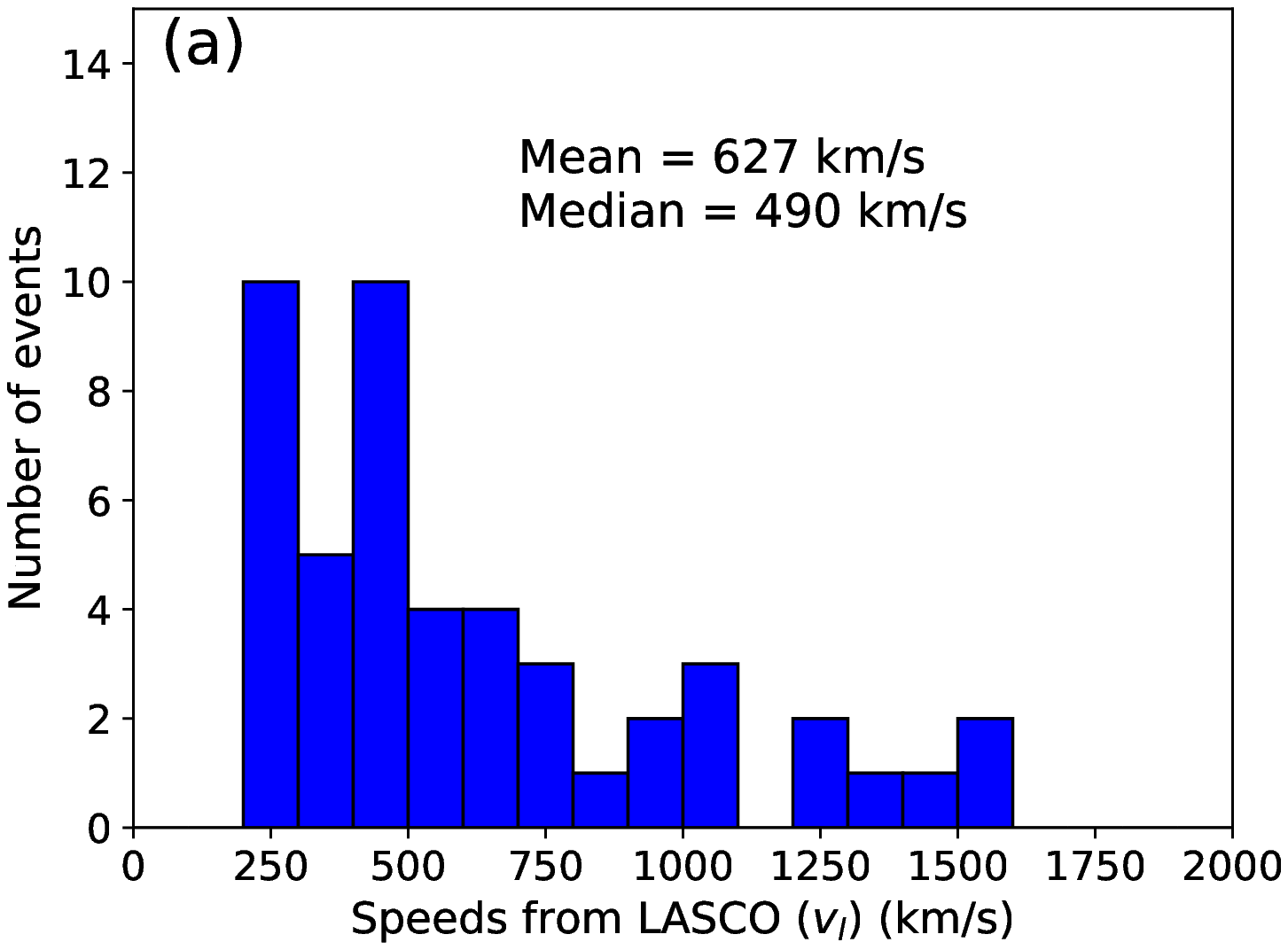}
\centering \includegraphics[width=0.49\textwidth]{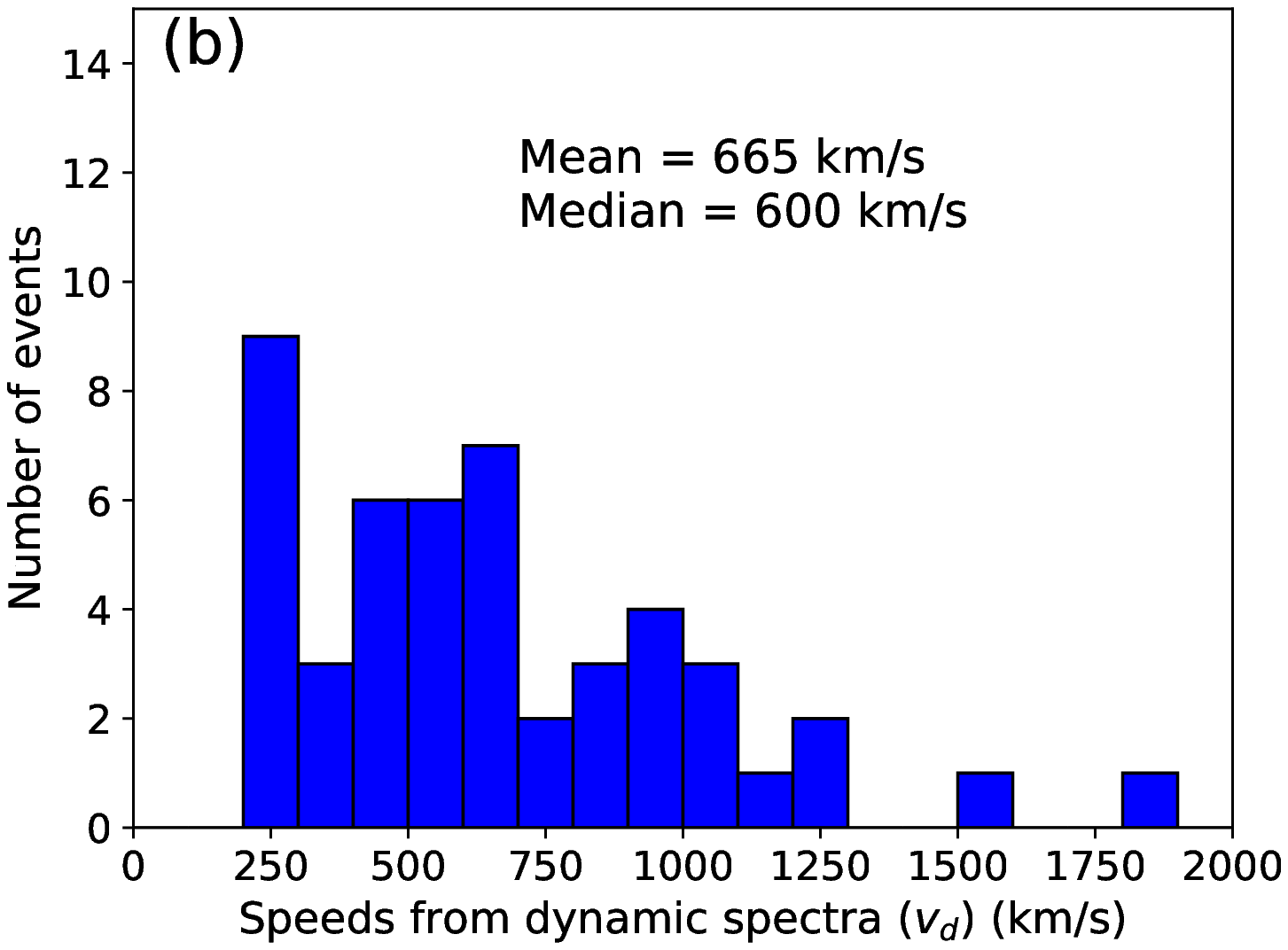}
\centering \includegraphics[width=0.49\textwidth]{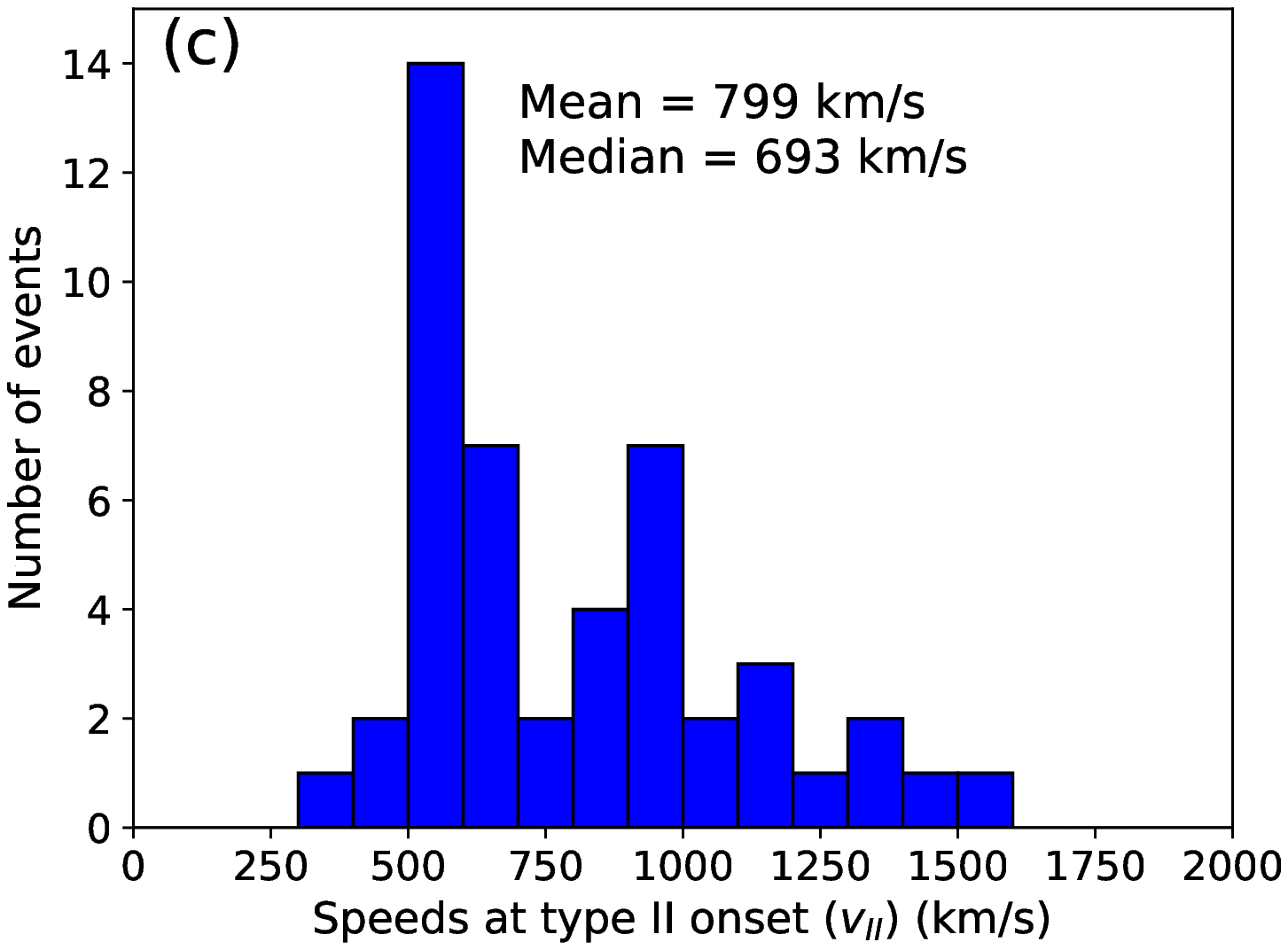}
\centering \includegraphics[width=0.49\textwidth]{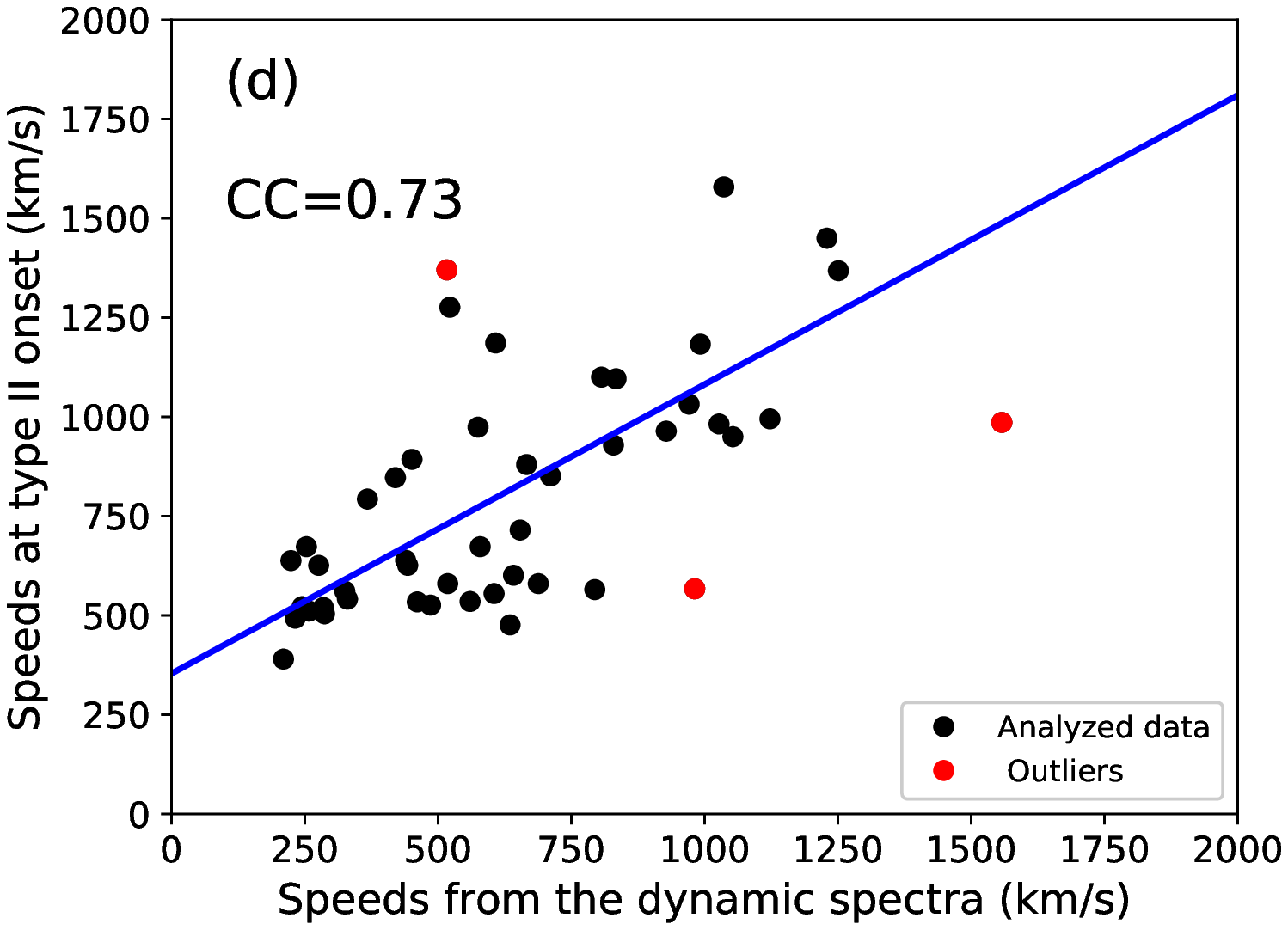}
\caption{(a) Distribution of CME speeds within LASCO FOV. The average and median CME speeds in the LASCO FOV are $\sim 627$ and $\sim 490$ km/s, respectively. (b) Distribution of CME speeds from the dynamic spectra, the observed average and median are $\sim 665$ and $\sim 600$ km/s. (c) Distribution of CME speeds at type II onset, the mean and median are $\sim 799$ and $\sim 693 $ km/s, respectively. The average CME speed at type II onset is higher compared to the speed in LASCO FOV, which is consistent with the shock formation. (d) The correlation between the average CME speed at type II onset ($v_{II}$) and the speed estimated using the corona density model ($v_{d}$) for the 48 events for which the shock could be measured. The correlation coefficient (cc) is $\sim 0.73$. By excluding the three outliers (in red), the cc improves to $\sim 0.82$. The `blue' line shows a linear fit to the data. 
}
\label{fig4}
\end{figure}

\noindent
Table \ref{table1} shows various parameters related to the HF type II bursts selected for this study and the associated CMEs and flares. The linear speed of the CME leading edge($v_l$) \citep{Yashiro2004, Gopalswamy09} observed with SOHO/LASCO is listed in column 5. The shock speed $v_d$ (in column 6) represents CME speed derived using the corona density model (see equation \ref{eq2})  from HF type II radio burst observations, and $v_{II}$ is the CME shock speed at type II onset obtained by linearly fitting the height-time measurements within STEREO FOV at the first appearance of type II bursts. \noindent Figure \ref{fig4}a shows the distribution of CME speeds within LASCO FOV. The average and median linear speeds of the CMEs are $\sim 627 $ and $\sim 490 $ km/s, respectively. Figure \ref{fig4}b shows the distribution of linear speeds estimated using parameters of the HF type II radio bursts spectra provided by CALLISTO. The average and median linear speeds of the CMEs are $\sim 665 $ and $\sim 600 $ km/s, respectively. The slightly higher linear speed estimates in the case of later are because this speed corresponds to the shock traveling ahead of the CME. Figure \ref{fig4}c is the distribution of CME shock speed at type II onset, the mean and median are $ \sim 799$ and $\sim 693$ km/s, respectively. The average speed of CMEs at type II bursts onset is higher than the average CME speed within LASCO and comparable with the speed derived using the HF type II burst parameters estimated from the dynamic spectra. This is consistent with the CME shock formation confirming a study by \cite{Gopalswamy2008} who found that the average speeds of CMEs associated with metric type II bursts are 610 km/s. Moreover, the determined average speed of the shocks ahead of the CMEs at type II radio bursts onset is very close to the shock speed, which excites metric type II bursts in the `lower' and `middle' corona (i.e., 1.1 - 3 $R_{\odot}$; \cite{Gosling1976}). Recently, \cite{Cunha-Silva} analyzed the properties of 4 type II bursts observed with CALLISTO spectrometer in Brazil and reported that the shock speed is in the range of 503 - 1259 km/s computed using Newkirk density model \citep{Newkirk}. \noindent Figure \ref{fig4} (d) shows the correlation between $v_d$ and $v_{II}$. The correlation coefficient (cc) is $\sim 0.73 $ and $\sim 0.79 $ by excluding the two outliers in red color. This high correlation coefficient values indicate that the speed derived from the coronal density model can be used as a proxy for estimating initial kinematics and dynamics of a CME in the lower corona in the absence of white-light observations \citep{kumari2017a, kumari2019direct}. The radio data can also be used as a proxy for disk events, for which estimating the initial speed of CMEs is difficult due to projection effects. 

\subsection{Case study}
\begin{figure}[h!]
\centering\includegraphics[width=6cm,height=5cm]{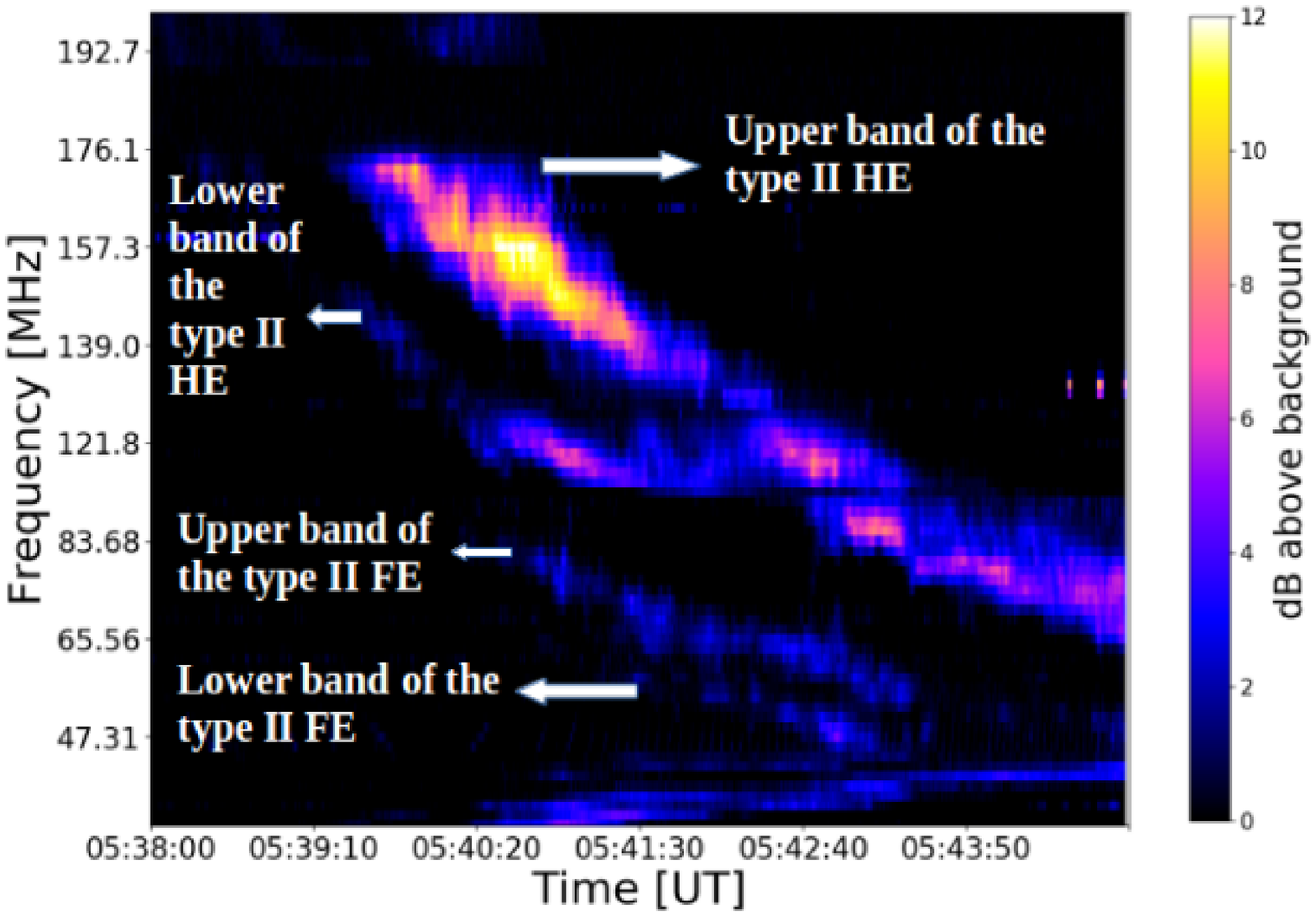}
\centering\includegraphics[width=5cm,height=5cm]{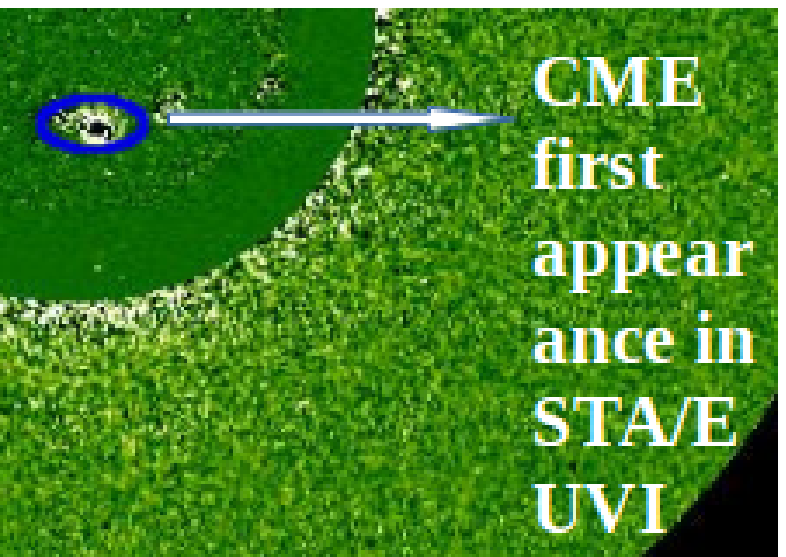}
\centering\includegraphics[width=5cm,height=5cm]{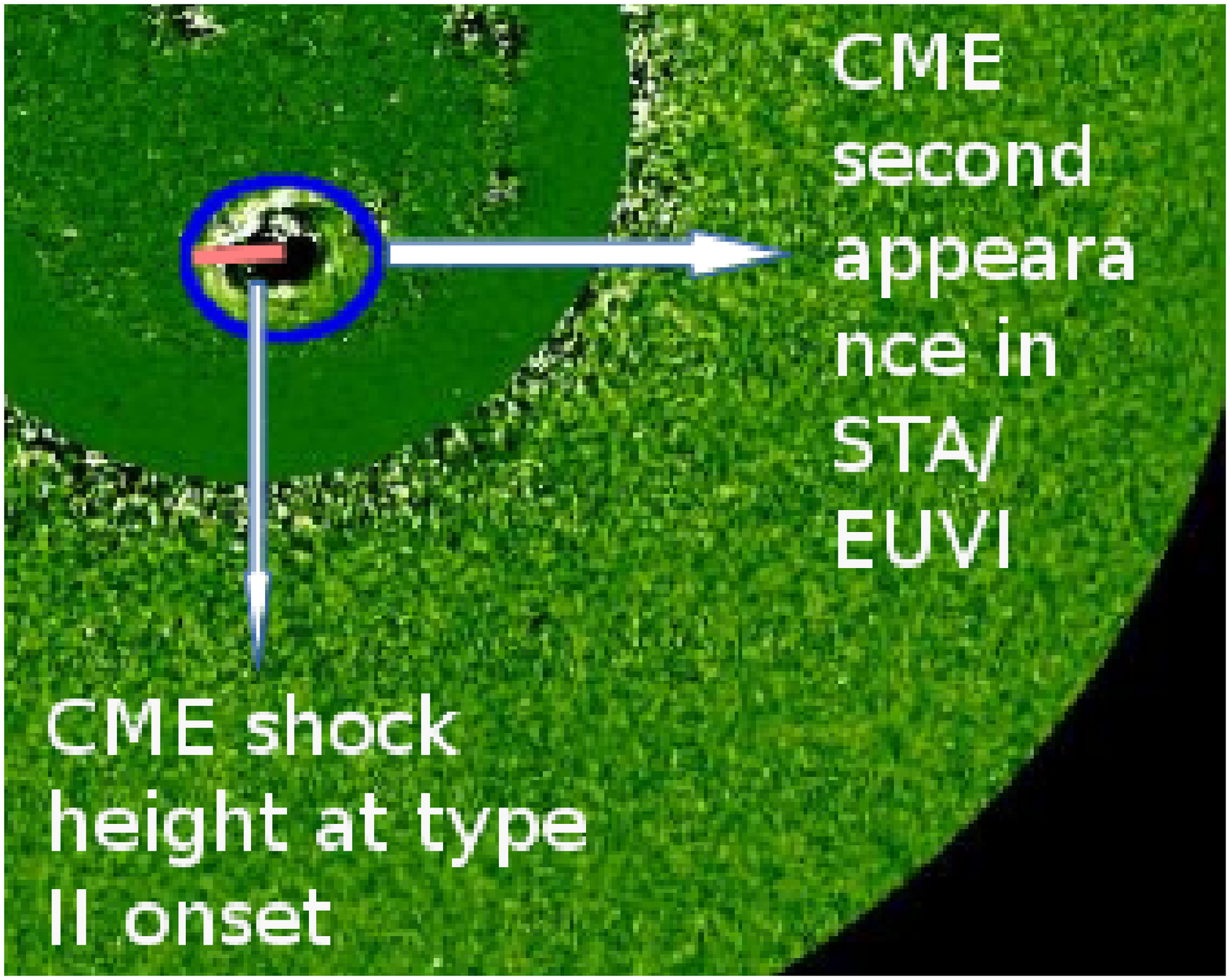}

\centering\includegraphics[width=6cm,height=5cm]{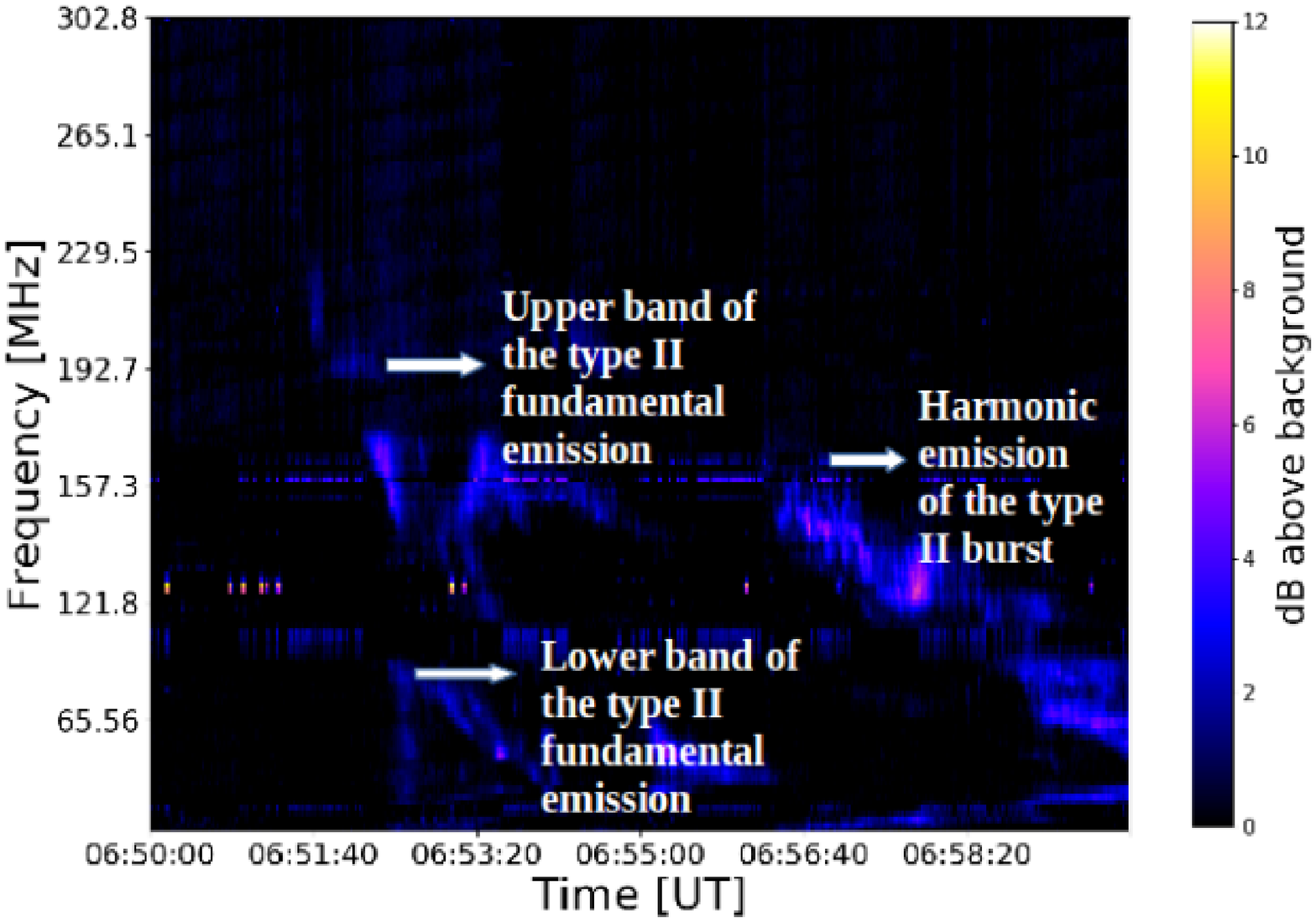}
\centering\includegraphics[width=5cm,height=5cm]{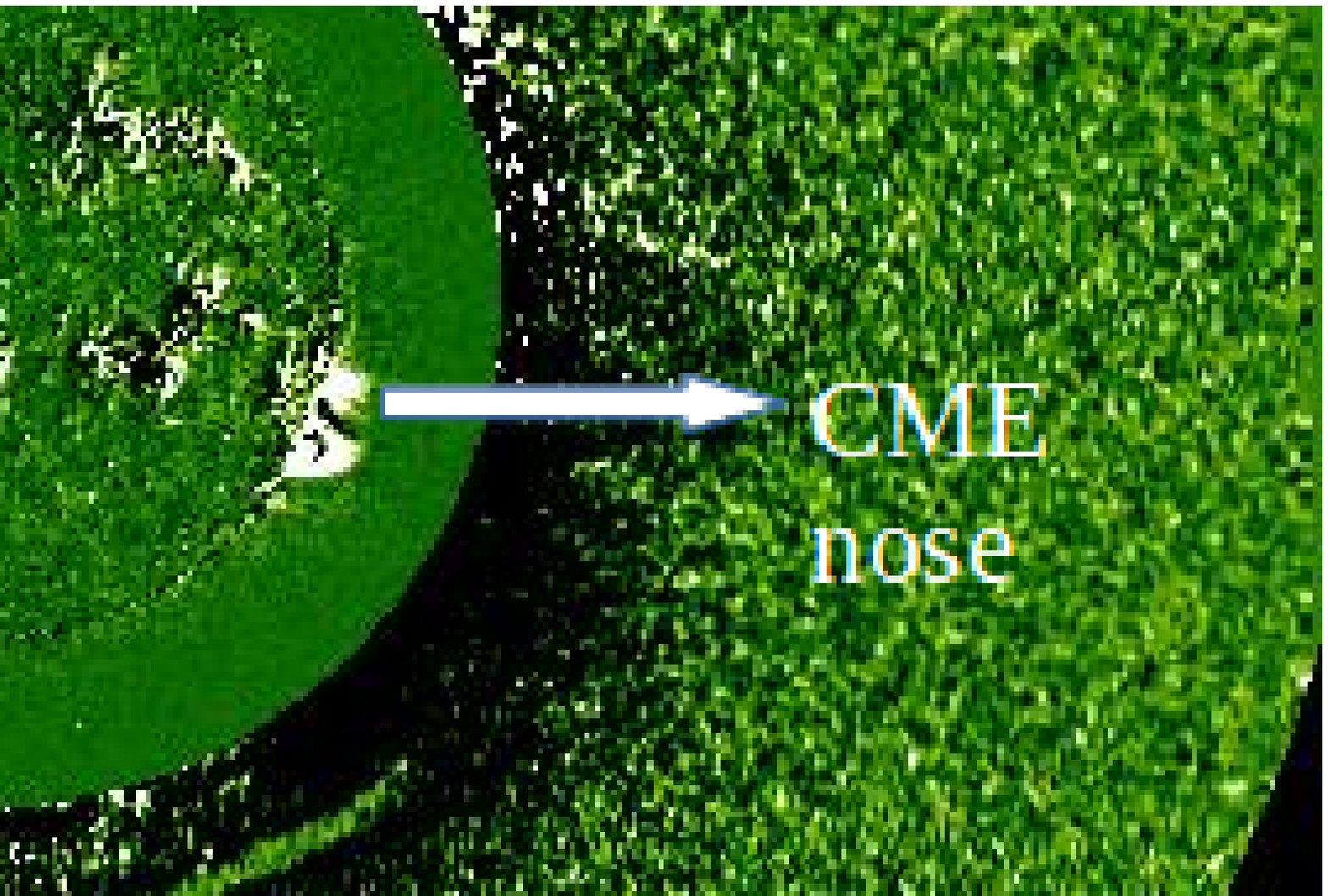}
\centering\includegraphics[width=5cm,height=5cm]{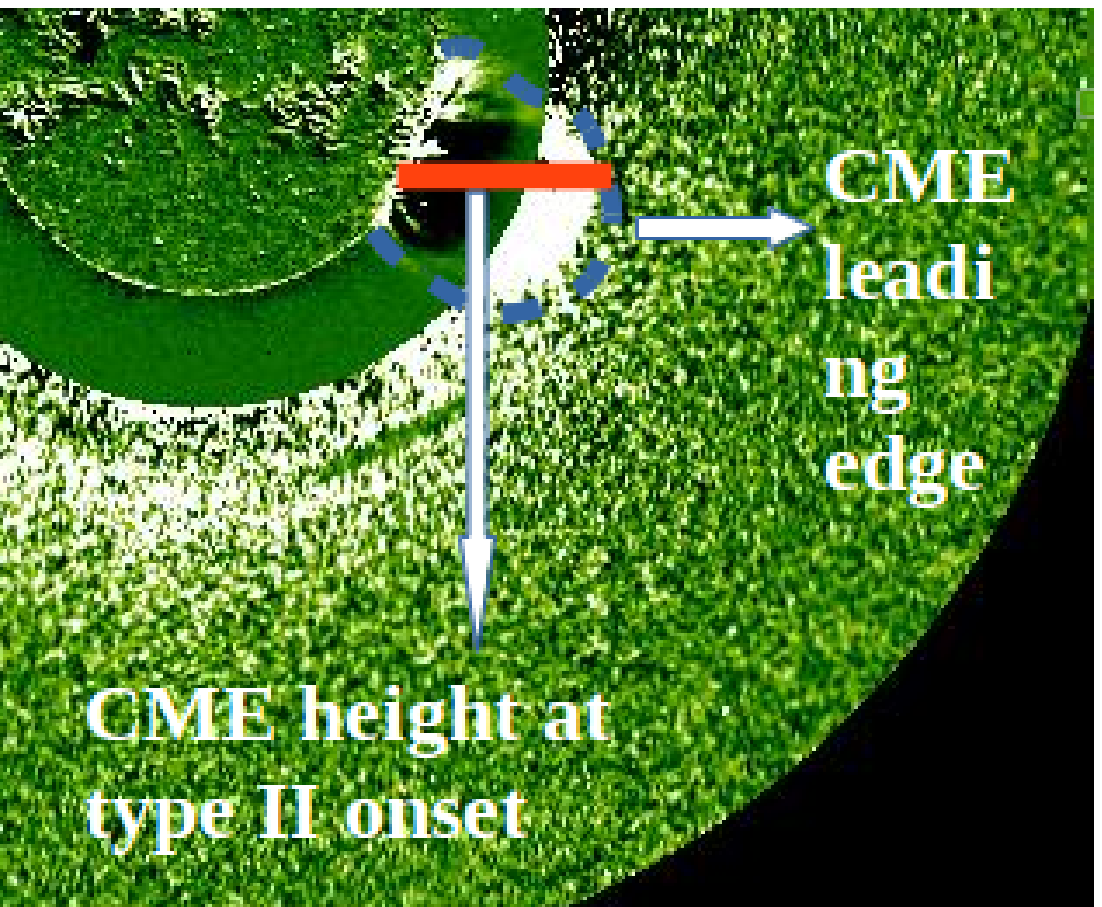}
\caption{\textbf{Top panel:} The dynamic spectra of HF type II burst observed with the with the CALLISTO spectrometer at Gauribidanur Radio Observatory on 13 June, 2010 (note that FE and HE stands for fundamental and harmonic emissions). The start frequency of this busts is $\sim 190$ MHz. Various features of this burst are marked on the spectra. The CME seen in STEREO FOV is also shown in this panel. The `blue' circles represent the fitting of a circle to the EUV wave to measure the shock formation height. 
\textbf{Bottom panel:} The dynamic spectra of HF type II burst observed with the CALLISTO spectrometer at GRO on 05 October, 2013. The start frequency of this busts is $\sim 240$ MHz. Various features of this burst are marked on the spectra. The CME seen in STEREO FOV is also shown in this panel.}   
\label{fig5}
\end{figure}

\noindent Event 1 (Type II burst observed on 13 June 2010): A HF type II burst was observed with the CALLISTO spectrometer at the Gauribidanur Radio Observatory (GRO)\footnote{\url{https://www.iiap.res.in/centers/radio}} on 13 June 2010 from 05:39-05:45 UT. Figure \ref{fig5} (top panel) shows the dynamic spectra of the type II burst. The start frequency (harmonic emission) of the burst is $\sim 190$ MHz. This type II burst had Fundamental, and Harmonic (F-H) bands and each of these two bands was further split into the split band. This radio burst had a drift rate of $\sim 0.28$ MHz/s. It was associated with a CME with a linear speed of $\sim 320$ km/s in SOHO/LASCO FOV. The estimated linear speed of the CME shock for the type II burst was found to be $\sim 375$ km/s. During the onset of the type II burst, the CME must have been at the radial distance of $\sim $ 1.17 Rs, measured using the wave diameter method (CME shock height is taken as the radius of the circle fitted to the outermost part of the disturbance, we assumed that the disturbance expands spherically above the solar surface. The radius is shown by a pink line on the last image frame of CME in Figure \ref{fig5}  top panel. More details about the method are found in \cite{Gopalswamy2013}). The type II burst was associated with an M1.0 class X-ray flare that originated at S23W75 on the solar disk\footnote{\url{https://www.solarmonitor.org/?date=20100613}}. Figure \ref{fig5} (top panel) also shows the CME in STEREO-A/EUVI and cor1 FOV. We used wave diameter method to estimate the CME shock height, where the CME shock height is taken as the radius of the sphere. Note that the event was detected by both STEREO and SDO spacecraft. We used STEREO/EUVI data for the height estimates and verified the same with SDO/AIA data. 

\noindent Event 2 (Type II burst observed on 05 October 2013): 
Another HF type II burst was observed with the CALLISTO spectrometer at the same radio observatory on 05 October 2013 from 06:52-07:13 UT. Figure \ref{fig5} (bottom panel) shows the dynamic spectra of the type II burst. The start frequency (harmonic emission) of the burst is $\sim 240$ MHz. This type II burst had complex FH split bands. This radio burst had a drift rate of $\sim 0.66$ MHz/s. 
This type II burst was associated with a CME with a linear speed of $\sim 964$ km/s in SOHO/LASCO FOV. The estimated linear speed of the CME shock for the type II burst was found to be $\sim 1006$ km/s. The measured CME shock height using the leading method \citep{Gopalswamy2013} as found to be $\sim $ 1.72 Rs at 06:55:00 UT (the red line shows the shock height in Figure \ref{fig5}, bottom panel, last CME image frame). The type II burst was associated with a C2.0 class X-ray flare that originated at S03E43 on the solar disk\footnote{\url{https://www.solarmonitor.org/?date=20131005}}. Figure \ref{fig5} (top panel) also shows the CME in STEREO-A/EUVI and cor1 FOV. We used the leading edge method as mentioned above to estimate the shock height. 

\section{Conclusion}
\label{sect.5}
\noindent
In this study, we analyzed 51 high starting frequency type II solar radio bursts events which occurred during 2010-2019 detected by CALLISTO spectrometers. Ground-based CALLISTO instruments in metric wavelengths could detect 51 out of 180  high frequency (ranging between 150-450 MHz) type II bursts. We carried out a statistical study to check the origin of the analyzed high-frequency type II bursts. The analyzed CMEs associated with the 51 HF type II bursts are wide with the average and median widths of $201^{\circ}$ and $175^{\circ}$, respectively.  We used the corona density model as defined in a study by \cite{Gopalswamy2006, Gopalswamy2011a} to convert the parameters obtained with radio observations into the shock speed. The type II parameters derived from the dynamic spectra, such as the drift rates, are used in the empirical relations to derive the CME's shock speed associated with these radio bursts. The mean and median CME/shock speeds in the LASCO FOV, radio observations, and type II onset were $\sim 627/665/799$ and $\sim 490/600/693$ km/s, respectively. At type II onset, the CME shock speed was high enough to drive a shock ahead of a CME.  In this work, the CME shock heights at type II onset were determined using either the leading edge method or wave diameter method depending on the CME location on the solar disk. The shock ahead of the analyzed CMEs was found within the heliocentric distances 1.17 to 2.02 Rs with a mean and median of 1.52 and 1.49 Rs. The time difference of 15 minutes between the first appearance of the CME in the STEREO and the observation of the fundamental band of the metric type II was considered to confirm their association. Note that it was difficult to distinguish the type II burst fundamental and harmonic emissions for some cases. Therefore, the study analyzed particularly type II bursts with the starting frequency $\geq$ to 300 MHz assuming harmonic emissions. The analysis found that all the analyzed HF type II bursts are due to  CMEs, hence confirming the previous studies.\\[0.5cm] \\
\noindent
Of the 51 type II bursts 45 were associated with solar flares. We found that the flares associated with high-frequency type IIs have shorter rise time and duration and are stronger and more impulsive. Also, the analysis shows that more flares associated with the analyzed type II bursts originate from the active regions with heliographic latitude $\pm 25^{\circ}$. Also, the number of the bursts from the western longitudes is larger than that from the eastern longitude. Using the CME properties associated with the analyzed high starting frequency bursts such as speed, width, and solar source longitude, we found that 9/45 events are Solar Energetic Particle (SEP)-accompanied and 7/9 originate from the western hemisphere. This could be due to the magnetic connectivity between the observer located at the Earth and SEP source region. Similar results were previously obtained by \cite{Gopalswamy2003, Gopalswamy2006, Gopalswamy2008} We also evaluated the performance of the coronal density model used to estimate the CMEs speed associated with type II bursts by comparing the estimated speed with the average CME shock speeds at type II onset using height-time measurements within STEREO FOV. The correlation is $\sim 73\%$.\\[0.5cm] \\
\noindent
As a case study, we analyzed two individual bursts with the Gauribidanur station of the e-Callisto network. Using the ground-based e-Callisto network with different longitudes, we can detect solar radio bursts within 24 hours a day, which serves as a warning of the associated solar transient, hence crucial for space weather studies \citep{Ndacyayisenga_2021}. This analysis agrees with previous similar studies that higher frequency type II bursts can act as a proxy to estimate this eruption's early kinematics and dynamics near the Sun. Notably, the present study demonstrates the importance of using higher frequency type II bursts observed from the ground to monitor space weather prediction.

\section*{Acknowledgements}
\noindent
The authors acknowledge the financial support from the Swedish International Development cooperation Agency (SIDA) through the International Science Program (ISP) to the University of Rwanda (UR-Swedish program) through the Rwanda Astrophysics, Space and Climate Science Research Group (RASCSRG). We thank the data centre of the e-Callisto network which is hosted by the FHNW, Institute for Data Science in Switzerland.  Our sincere thanks goes to the providers of  SOHO/LASCO-c2 CAW list and SWPC event list. One of the authors, A.K. acknowledges the ERC under the European Union's Horizon 2020 Research and Innovation Programme Project SolMAG 724391.


\end{document}